\documentclass[sigconf]{acmart}
\pdfoutput=1
\usepackage[utf8]{inputenc}

\usepackage{algorithmic}
\usepackage{comment}
\usepackage{cleveref}
\usepackage{graphicx}
\usepackage{siunitx}
\usepackage{subfigure}
\usepackage{textcomp}
\usepackage{todonotes}
\usepackage{xcolor}
\usepackage{xspace}

\theoremstyle{definition}
\newtheorem{researchquestion}{RQ}
\crefname{researchquestion}{RQ}{RQs}

\newcommand{\SWH}{Software Heritage\xspace}

\newcommand{\CommitCount}{2.2 billion\xspace}
\newcommand{\AuthorCount}{43 million\xspace}
\newcommand{\ProjectCount}{160 million\xspace}

\def\DATAAuthorsRaw/{\num{43 381 366}}
\def\DATAAuthorsPlausible/{\num{33 351 300}}
\def\DATACommitsRaw/{\num{2 198 808 389}}
\def\DATACommitsPlausible/{\num{1 735 130 408}}
\def\DATATotalCommitsInSH/{about \ProjectCount}
\def\DATALastCommitDate/{2021-07-07}
\def\ALGtzname/{tz\&name}
\def\ALGemail/{ccTLD}
\def\DATACommitsWithEmail/{13\%}
\def\DATACommitsWithoutEmail/{87\%}
\def\DATACommitsTZZTwoThousand/{96\%}
\def\DATACommitsTZZTwoThousandTen/{64\%}
\def\DATACommitsTZZTwoThousandTwenty/{22\%}
\def\DATAWorldRegions/{12}  \def\PKGGG{\mbox{\scshape\small gender-guesser}\xspace}
\def\TZANALYZED/{20}
\def\DATAZonedCommitsRatio/{70\%}

\def\DATAKnownAuthorsPct/{64.1\%}
\def\DATAUnknownAuthorsPct/{35.9\%}
\def\DATAKnownAuthorsApprox/{21.4\,M}
\def\DATAKnownCommitsPct/{50.3\%}
\def\DATAUnknownCommitsPct/{49.7\%}
\def\DATAKnownCommitsApprox/{872.3\,M}

\def\DATAMaleAuthorsPct/{86.5\%}
\def\DATAMaleAuthorsApprox/{10.4\,M}
\def\DATAFemaleAuthorsPct/{13.5\%}
\def\DATAFemaleAuthorsApprox/{1.6\,M}

\def\DATAMaleCommitsPct/{91.9\%}
\def\DATAMaleCommitsApprox/{801.8\,M}
\def\DATAFemaleCommitsPct/{8.1\%}
\def\DATAFemaleCommitsApprox/{70.5\,M}

 \copyrightyear{2022}
\acmYear{2022}
\setcopyright{acmlicensed}\acmConference[ICSE-SEIS'22]{Software Engineering in
Society}{May 21--29, 2022}{Pittsburgh, PA, USA}
\acmBooktitle{Software Engineering in Society (ICSE-SEIS'22), May 21--29,
2022, Pittsburgh, PA, USA}
\acmPrice{15.00}
\acmDOI{10.1145/3510458.3513011}
\acmISBN{978-1-4503-9227-3/22/05}

\title{Worldwide Gender Differences in Public Code Contributions}
\subtitle{and how they have been affected by the COVID-19 pandemic}
\author{Davide Rossi}
\email{daviderossi@unibo.it}
\affiliation{\institution{University of Bologna}
  \city{Bologna}
  \country{Italy}
}

\author{Stefano Zacchiroli}
\email{stefano.zacchiroli@telecom-paris.fr}
\orcid{0000-0002-4576-136X}
\affiliation{\institution{LTCI, Télécom Paris, Institut Polytechnique de Paris}
  \city{Paris}
  \country{France}
}

 \begin{abstract}
  Gender imbalance is a well-known phenomenon observed throughout sciences
  which is particularly severe in software development and Free/Open Source
  Software communities. Little is know yet about the geography of this
  phenomenon in particular when considering large scales for both its time and
  space dimensions.

  We contribute to fill this gap with a longitudinal study of the population of
  contributors to publicly available software source code. We analyze the
  development history of \ProjectCount software projects for a total of
  \CommitCount commits contributed by \AuthorCount distinct authors over a
  period of 50 years. We classify author names by gender using name frequencies
  and author geographical locations using heuristics based on email addresses
  and time zones. We study the evolution over time of contributions to public
  code by gender and by world region.

  For the world overall, we confirm previous findings about the low but
  steadily increasing ratio of contributions by female authors. When breaking
  down by world regions we find that the long-term growth of female
  participation is a world-wide phenomenon. We also observe a decrease in the
  ratio of female participation during the COVID-19 pandemic, suggesting that
  women's ability to contribute to public code has been more hindered than that
  of men.
\end{abstract}

\keywords{gender, diversity, open source, commit, software heritage, covid19}

\begin{document}
\maketitle

\section*{Lay Abstract}
\label{sec:layabstract}

Software developers around the world work together to produce publicly available software (or \emph{public code}).
They do so using public identities and disclosing information about their work that include their names and when a software change was made.
We use this information to characterize the gender gap in public code, that is, the difference in participation to public software development between men and women.
Specifically, we study the development history of \ProjectCount pieces of public software, developed over a period of 50 years by \AuthorCount authors.
We characterize the gender gap on this corpus over time and by world region.
To determine author genders we rely on public data about name frequencies by gender around the world.
To determine author locations we use email addresses, name frequencies around the world, and the timezone associated to each software change.
We confirm that the gender gap in public code is huge.
Female authors are only \DATAFemaleCommitsPct/ of the total and have authored only \DATAFemaleAuthorsPct/ software versions.
The gender gap is however shrinking, with women participation having increased steadily over the past 12 years.
This improvement is a global phenomenon, observable in most world regions.
We also observe a decrease in the ratio of female participation during the COVID-19 pandemic, suggesting that women have been more hindered than men in their ability to contribute to public code.

 \section{Introduction}
\label{sec:intro}

\emph{Gender imbalance} (or \emph{gender gap}) is the situation in which,
within a given group of people in society, a gender is significantly over- or
under-represented with respect to the gender partition that exists in the world
at large, which is close to a 50\%/50\% men/women split~\cite{un2019worldpop}.
Gender imbalance \emph{tilted toward women under-representation} and men
over-representation in academia has been observed in several fields and is
particularly severe across STEM disciplines (science, technology, engineering,
and mathematics)~\cite{hill2010whysofew, wang2017genderstem,
  unesco2017genderstem, reinking2018genderstem, botella2019genderstem}. This
gender gap is even more acute in computing, where it has been so for many
decades now~\cite{margolis2002womencs, frenkel1990womencs}, with only recent
signs of being on the decline~\cite{zacchiroli2021gender} and a long way to go
before reaching gender parity.

In the context of software development, Free and Open Source Software (FOSS)
projects have been frequently analyzed from the gender imbalance angle,
confirming multiple times~\cite{david2008fossdevs, qiu2010kdewomen,
  kuechler2012genderfoss, vasilescu2014gender, vasilescu2015gender,
  oneil2016debiansurvey, robles2016womeninfoss, terrell2017gender,
  bosu2019fossdiversity, zacchiroli2021gender} the under-representation and very low participation (in
relative terms) of women in FOSS. These results have been obtained via
different techniques---from surveys to interviews and name-based analyses---and
at very different scales---from individual projects and focus groups to
very-large scale analyses of public code---leaving little to no doubts about
the existence of a gender gap in FOSS and public software development.

The \emph{geography of the gender gap in public code} is a relatively
under-explored angle of gender imbalance, which we explore in this paper at
very large scale and along two orthogonal dimensions: time and space. Along the
time axis, and following in the steps of recent related
work~\cite{zacchiroli2021gender}, we analyze more than 50 years of public code
contributions from the Software Heritage archive~\cite{swhipres2017},
consisting of more than \CommitCount commits contributed by \AuthorCount
distinct authors, and we classify them by gender using a frequency-based
approach applied to author names. 
This gives us a baseline \emph{worldwide
  historical trend of the evolution of the gender gap} in public code
contributions, which generally confirms previous results: the overall amount of
contributions authored by female authors is very low, but is also steadily
growing.

Along the space axis we break down code contributions geographically to verify
if there are significant differences in gender gap trends across different
world regions. Specifically, we will answer the following research questions:
\begin{researchquestion}
  \label{rq:gendertimezone} What is the overall breakdown \emph{by gender and
    UTC offset} in contributions (and contributors) to public source code?
\end{researchquestion}
\begin{researchquestion}
  \label{rq:genderworld} What is the overall breakdown \emph{by gender and by
    world regions} in contributions (and contributors) to public source code?
\end{researchquestion}

\cref{rq:gendertimezone} is our first approximation of the geographic position
of contributors worldwide. The UTC offset is the difference in minutes between
Coordinated Universal Time (UTC) and local time at a particular place on Earth.
As UTC offset values are spread along the East-West axis of the planet, and
following previous work~\cite{barahona2016geodistrib}, we use them to group
developers by (coarse-grained) longitude, before breaking down each group by
author gender.

\cref{rq:genderworld} is a more refined approximation of the location of commit
authors, this time at a granularity of a division of the world in \DATAWorldRegions/
regions, loosely based on the world sub-regions of the United Nations
geoscheme~\cite{un1999geoscheme}. To geolocate public code contributions and
their authors at this granularity we use a heuristic based on commit timezones,
name frequencies world-wide, and country code top-level domain (ccTLD) found in
commit emails.

In answering these two research questions we find that the long-term growth of
female participation is a world-wide phenomenon that is observable across all
UTC offsets and all zones in the analyzed corpus albeit with trends 
showing appreciable regional differences.

While analyzing the data used to answer
\cref{rq:gendertimezone,rq:genderworld}, we noticed a worldwide phenomenon: the
decrease of the ratio of women participation to the production of public code
for the year 2020---the year when the COVID-19 pandemic struck. Hence, although
that was not part of the initial study design, we also separately studied this
phenomenon and address it in the paper as the following research question:
\begin{researchquestion}
  \label{rq:gendercovid} Has the impact of the 2020 \emph{COVID-19 pandemic} on
  contributions to public code been quantitatively different by gender?
\end{researchquestion}

The historical trends, both world-wide and region-by-region, that emerge from
our analyses suggest that women have been more negatively hindered in their
ability to contribute to public code during the pandemic than men, reverting
for the first time in 2020 the positive trend of the ratio of contributions by
female authors, which had been growing steadily since the 90s.

\paragraph{Paper structure}
We review relevant related work in \cref{sec:related}. We detail our analysis
methodology in \cref{sec:methodology}. We present the obtained raw results and
discuss them in light of our research questions in \cref{sec:results}. Before
concluding, we discuss limitations and threats to validity in
\cref{sec:threats}.

\paragraph{Data availability}
A replication package for this paper is available from
Zenodo~\cite{replication-package}.

 \section{Related Work}
\label{sec:related}

In early work on this topic, Hill et al.~\cite{hill2010whysofew} summarized the
under-representation of women in STEM, documenting the quantitative extent of
the phenomenon and discussing when female students drop out from an initially
well-balanced funnel of students. More recent work~\cite{wang2017genderstem,
  unesco2017genderstem, reinking2018genderstem, botella2019genderstem} review
the \emph{status quo} for STEM, including quantitative and qualitative analyses
of the phenomenon, as well as analyses of which practices in society (including
in education) contribute to it.

Margolis and Fisher~\cite{margolis2002womencs} looked into the case of computer
science students, at college level and below, characterizing the gender gap in
the field and also exploring one of its main causes (i.e., the widespread and
self-reinforcing assumption that computing is a ``boys' clubhouse'') based on
interviews with a pool of more than 200 college students.

From a theoretical framework point of view, the gender gap in FOSS has been
explained by Nafus~\cite{nafus2012patches} as a consequence of interaction
practices that hinder women's inclusion. Empirically, the rapid increase of
free/open source software has attracted since the early 2000s' attention to the
gender gap in FOSS, which has been verified to be more severe than in computing
at large.

Researchers have resorted to different techniques to characterize the FOSS
gender gap. Multiple survey-based studies of FOSS contributors have reported
low ratios of women respondents. Surveys~\cite{david2008fossdevs} up to 2003
reported that 95--99\% FOSS participants self-identified as men. A more recent
survey of \num{2000} FOSS contributors in 2013~\cite{robles2016womeninfoss}
reported a ratio of 10\% women respondents. In addition to surveys who invited
FOSS contributors at large without preselecting projects, the FOSS gender gap
has been verified also within specific FOSS communities, such as
Debian~\cite{oneil2016debiansurvey}, KDE~\cite{qiu2010kdewomen}, and
OpenStack~\cite{izquierdo2019openstackdiversity}.

Specific artifacts resulting from software development activities, in both FOSS
and collaborative software development in general, have been analyzed to
quantify the gender gap. Mailing lists of early FOSS projects have been
analyzed for gender differences by Kuechler et
al.~\cite{kuechler2012genderfoss}, finding evidence of declining female
participation over time. Stack Overflow and GitHub teams have been studied for
analogous reasons by Vasilescu et al.~\cite{vasilescu2014gender,
  vasilescu2015gender}, finding compatible evidence. Terrell et
al.~\cite{terrell2017gender} studied pull requests on GitHub, finding evidence
of gender bias against code contributions coming from women outsiders. Bosu and
Sultana~\cite{bosu2019fossdiversity} mined code reviews of ten popular FOSS
projects, determining that only 10\% of active contributors were women.

Trinkenreich has designed~\cite{trinkenreich2021pleasedontgo} a holistic
research agenda to investigate how FOSS communities can actively increase the
participation of women in their projects. In preliminary work along that path
Trinkenreich et al.~\cite{trinkenreich2020hiddenfigures} identified different
career paths in FOSS, as each of them might require different engagement
strategies to attract and retain diverse contributors.  Canedo et
al.~\cite{canedo2020womencoredev} conducted a mixed-methods study to
investigate the gender gap for women \emph{core developers} in FOSS, confirming
its existence but noting that interviewed women core developers did not report
having experienced gender discrimination in FOSS.

Rastogi~\cite{rastogi2016geobias} has studied a different form of bias in
public code development, that related to geographical location, showing that it
has an impact on maintainers' decisions on whether to accept a pull request or
not. Lacking pull request data in our dataset it is not something we can
explore, but we provide baseline data about developer origins that can support
future similar studies.

A very large scale gender study of public code contributions has been conducted
by Zacchiroli~\cite{zacchiroli2021gender}, studying all public code commits
available from Software Heritage to observe the evolution over time of the
gender gap. In this study we use the same data source and the same tool
(\PKGGG) for gender identification, with the following differences: (1) we
characterize the geography of the FOSS gender gap, allowing comparisons across
world regions; (2) we use a more recent data set, covering 1 more year
(specifically: 2020, allowing to answer \cref{rq:gendercovid}).

The impact of the COVID-19 pandemic on the gender gap due to an uneven
distribution of caregiver responsibilities among parents has been investigated
in several studies~\cite{alon2020covidgender, collins2021covidgender}. Ralph et
al.~\cite{ralph2020pandemic} have investigated the effects of the COVID-19
pandemic on software development specifically, via a survey of more that
\num{2000} developers worldwide. They tentatively suggest that \emph{``women,
  parents and people with disabilities may be disproportionately
  affected''}. In answering \cref{rq:gendercovid} we provide the first body of
empirical evidence that the pandemic has indeed disproportionately affected
women's ability to contribute to public code.

Across several works~\cite{robles2006devlocation, barahona2008geography,
  barahona2016geodistrib} Barahona et al.~have established a methodology to
pinpoint (coarse-grained) developer locations using various signals that
include commit timezones, email domains, and mailing list participation. We
adopt similar heuristic, using as signal only information coming from public
code commits (namely: names, timezones and email ccTLDs).

A very recent study by Prana et al.~\cite{prana2021geogenderdiversity} (``to
appear'' at the time of writing) pursues similar objectives to ours, and
\cref{rq:genderworld} specifically, albeit with a different approach. The
authors also study the gender gap in FOSS worldwide, but do so using a
mixed-methods approach consisting of repository mining and targeted developer
surveys. The scale of the present study is much larger both in terms of
analyzed repositories (160\,M v.~22\,K) and time period (50 years v.~7); the
drawback of this scale is that we could not further corroborate our findings
with data coming from developers via surveys. In terms of findings we observe
the same general trends of worldwide-low gender diversity and fast(er) increase
in specific world regions, although details differ slightly across regions.

 \section{Methodology}
\label{sec:methodology}

\begin{figure*}
  \centering
  \includegraphics[width=1.0\linewidth]{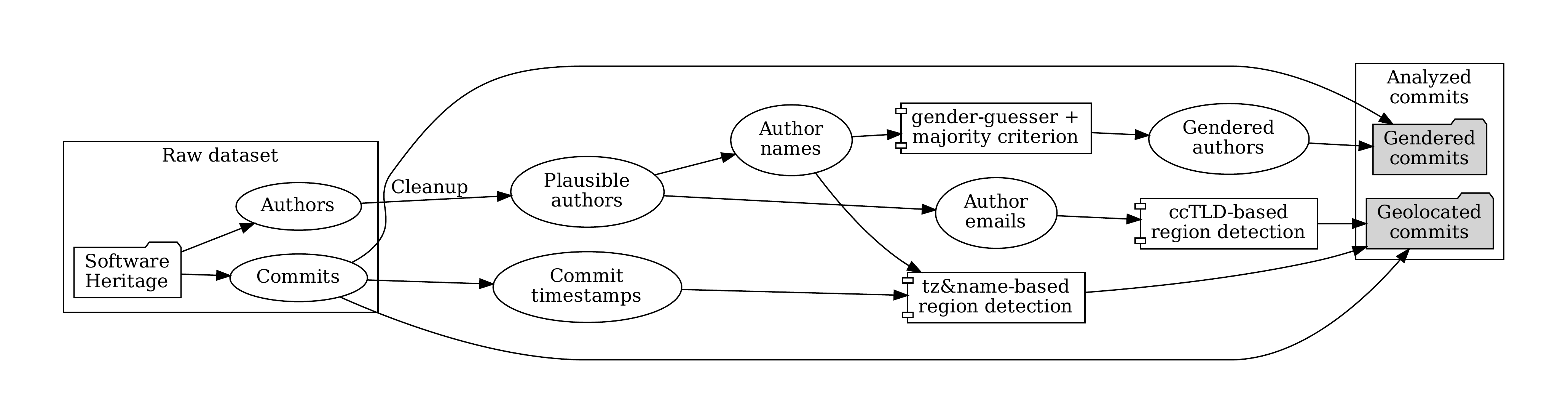}
  \caption{
    Analysis data flow.
    Starting from the Software Heritage dataset we detect the gender of commit authors using \PKGGG on name tokens and a majority criterion.
    We geolocate commits with two methods: (1) using country-level top-level email domains (e.g., \texttt{.uk}) extracted from commits; and (2) comparing commit author names to the most popular names among the countries that have a timezone compatible with the commit timestamp offset.
  }
  \label{fig:methodology}
\end{figure*}

The data flow and main components of the methodology adopted to answer \cref{rq:gendertimezone,rq:genderworld,rq:gendercovid} are depicted in \cref{fig:methodology}.

\paragraph{Terminology considerations}

To answer our research questions we need to assign a gender and a world region to the authors of commits in the dataset.
Some terminology considerations are in order about both axes.

Regarding gender, in the following we will refer to automated classification decisions described as ``gender detection'' or ``gender assignment''.
With that we do not intend to arbitrarily define people within a binary gender confinement regardless of their preferences and sensitivity.
None of the gender-related decisions made by the automated techniques used in this paper make sense when applied to \emph{individuals} included in the analyzed corpus.
The meaning of the exercise is statistical in nature and aims only to address the stated research questions.
The used approach makes sense only in aggregate form and carries with it the unavoidable limitations that name-based gender detection entail; we elaborate on those limitations in \cref{sec:threats}.

\begin{figure}
  \centering
  \includegraphics[clip,trim=6cm 6cm 0 0,width=\linewidth]{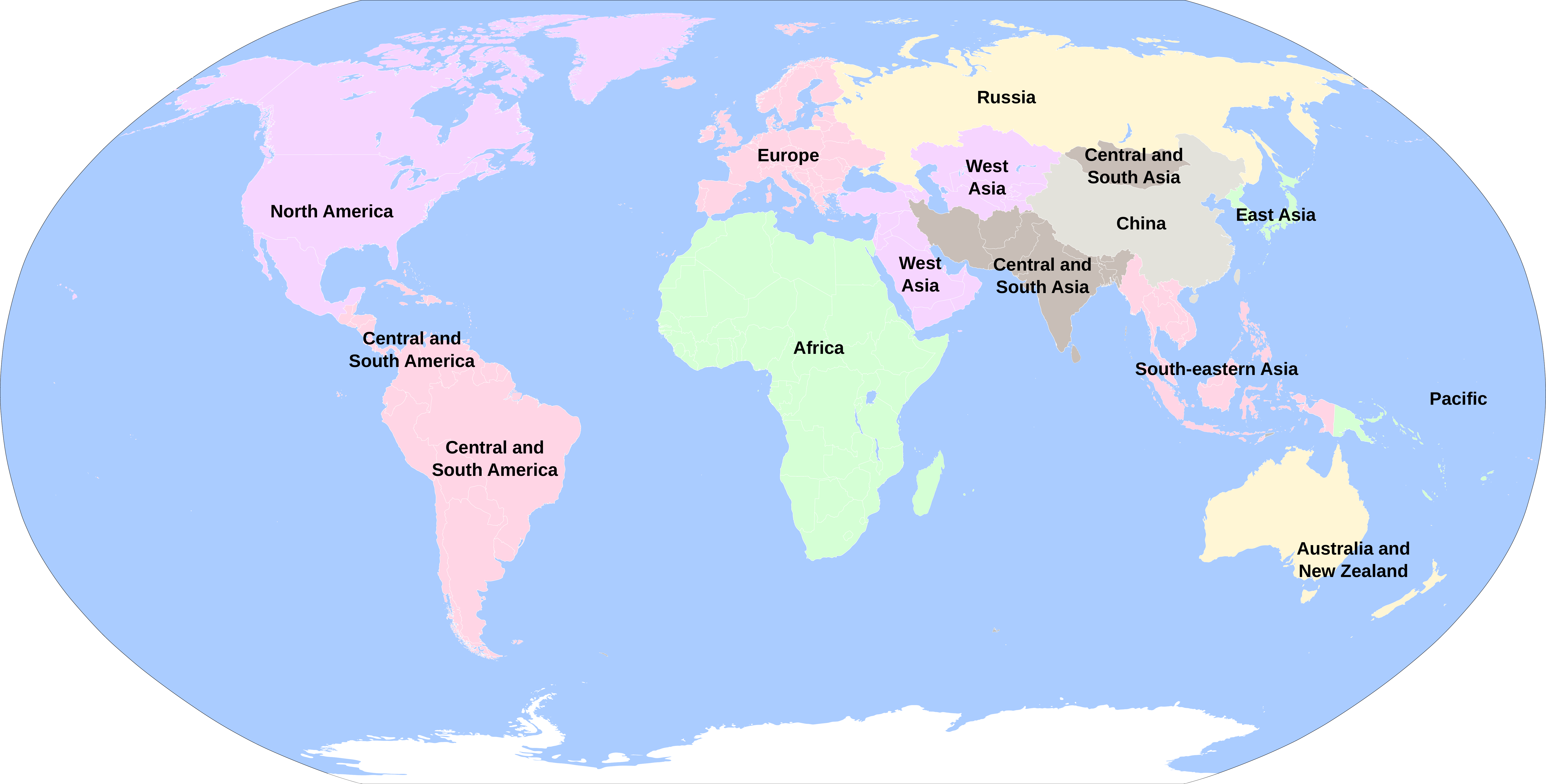}
  \caption{
    Map of the \DATAWorldRegions/ world regions used in this study to geolocate public code contributions. }
  \label{fig:worldmap}
\end{figure}

Regarding the geolocation of commits and their authors, we only consider the macro geographical areas in which contributions are likely to have been made, at the granularity of large world regions.
As geolocation targets we use the \DATAWorldRegions/ regions shown in \cref{fig:worldmap}, namely (in alphabetical order): Africa, Australia and New Zealand, Central and South America, Central and South Asia, China, East Asia, Europe, North America, Pacific, Russia, South-eastern Asia, West Asia.
To obtain them we started from the United Nations geoscheme~\cite{un1999geoscheme} as devised by the United Nations Statistics Division, to which we applied some merges and split based on geographical distance and/or on the sharing of preeminent cultural identification features, such as spoken language, only when needed to avoid under- or over-represented data samples (e.g., to avoid that China dominates the East Asia subsample or Russia the East Europe one).
Other than that, considerations similar to those made about gender also apply to national or cultural identities: no assessment about the identity of individuals in the corpus is implied, obtained figures only make sense in aggregate form, and limitations (discussed in \cref{sec:threats}) apply.

\subsection{Dataset}

As starting point we retrieved from \SWH~\cite{swh-msr2019-dataset} a snapshot of all the commits archived until \DATALastCommitDate/.
It consists of \DATACommitsRaw/ commits, unique by SHA1 identifier, harvested from \DATATotalCommitsInSH/ public projects coming from major development forges (GitHub, GitLab, etc.) and source code distributions (Debian, PyPI, NPM, NixOS, etc.).
Commits in the dataset have been contributed by \DATAAuthorsRaw/ authors, unique by $\langle$name, email$\rangle$ pairs.

Obtained commits came as two relational tables, one for commits and one for authors, with the former referencing the latter via a foreign key.
Each row in the commit table contains the following fields: commit SHA1 identifier, author and committer timestamps, author and committer identifiers (referencing the author table).
The distinction between commit authors and committers come from Git, which allows to commit a change authored by someone else.
For this study we focused on authors and ignored committers, as the difference between the two is not relevant for our research questions and the amount of commits with a committer other than its author is negligible.
For each entry in the author table we have author full name and email as two separate strings of raw bytes.

Looking into the raw author full names we realized that some of them are not real author names, but rather emails or gibberish strings, likely coming from misconfigured VCS tools.
Hence as a preliminary analysis steps we filtered out implausible or unusable author names in the dataset, such as: names that cannot be decoded as UTF-8 strings, email addresses used as names, names consisting of only blank characters, names containing more than 10\% non-letters, and names longer than 100 characters.
We did not perform any other data filtering or selection.
After filtering, \DATAAuthorsPlausible/ ``plausible'' authors remained (76.9\% of the initial authors), having authored \DATACommitsPlausible/ commits (78.9\% of the initial commits).

\begin{figure}
  \centering
  \def\growthwidth{0.9\linewidth}
  \includegraphics[width=\growthwidth]{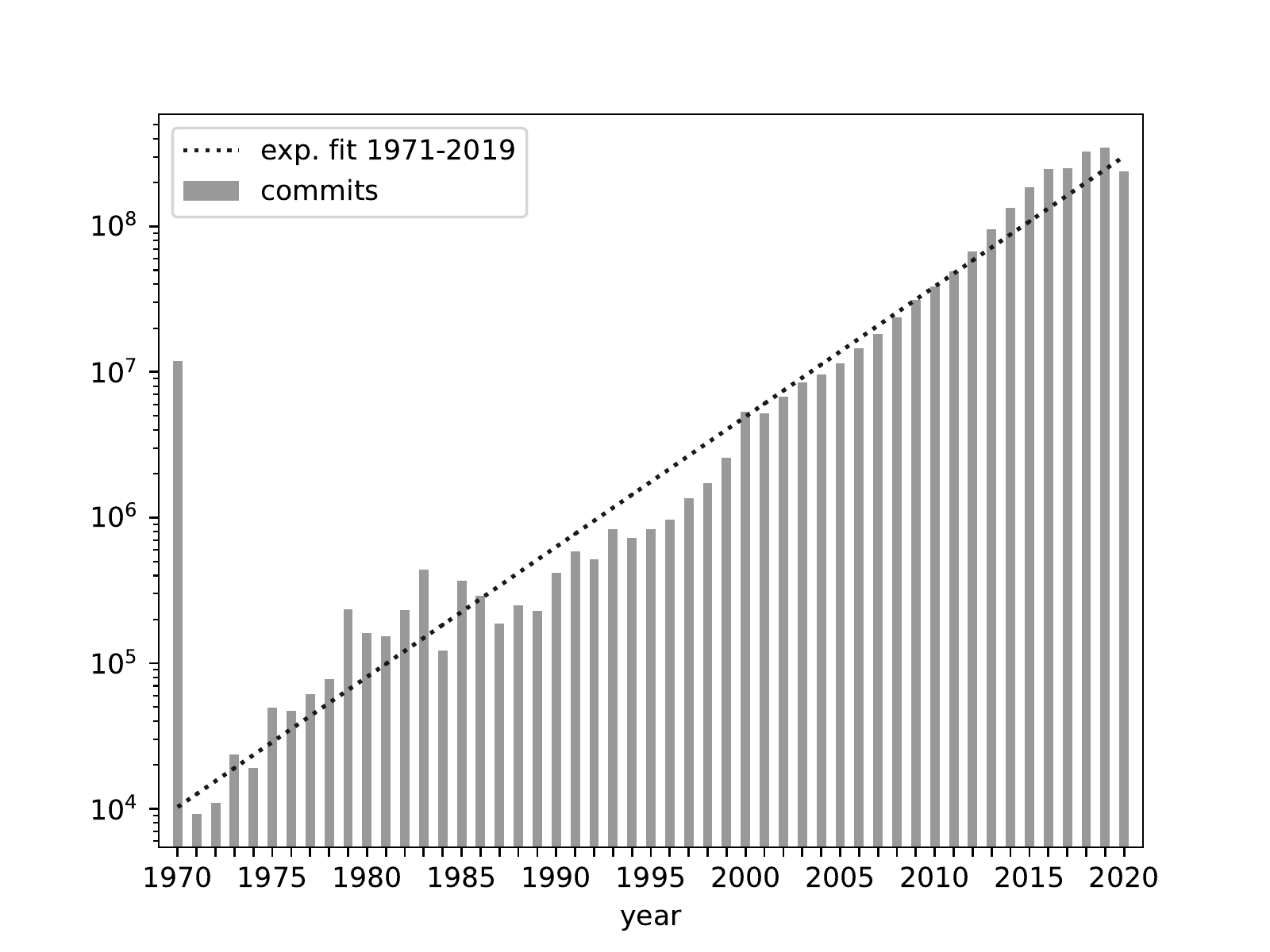}\\
  \includegraphics[width=\growthwidth]{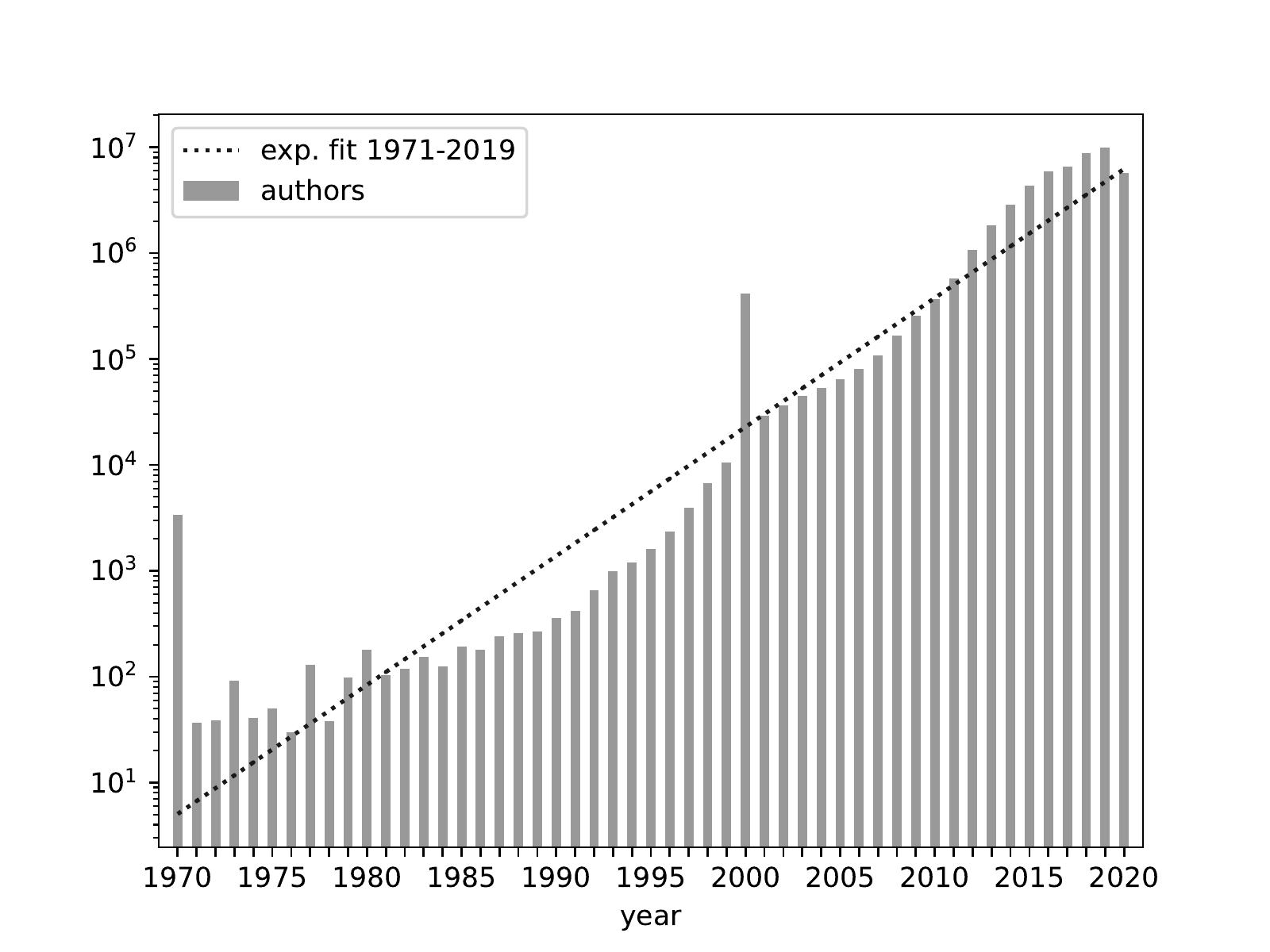}
  \caption{Number of commits (top) and authors (bottom) over time, with exponential fit for the 1971--2019 period.
    (Note the log-scale on the Y-axis.)
}
  \label{fig:growth}
\end{figure}

\Cref{fig:growth} shows the evolution over time of the commits in the dataset, for the period 1970--2020\footnotemark.
\footnotetext{We have restricted our analyses to commits with author timestamps in the 1970--2020 range.
  A limited amount ($<3\%$) of commits with timestamps outside that range exist in the dataset, partly due to when the dataset was obtained from \SWH (March 2021), partly due to the use of Git to model the history of historical documents such as the U.S.~Constitution, and partly due to misconfigured VCS tools resulting in author timestamps in the future.}
The dataset appears to grow exponentially over time, both in terms of commits and authors in it, and has done so for almost 50 years now.
Exceptions are the first and last (complete) years in the dataset, respectively 1970 and 2020, for different reasons.
1970 contains the UNIX epoch (1 January 1970 at midnight UTC), which is often used as ``default'' timestamp for older or missing points in time, ending up being over-represented.
2020 being close to when we obtained our dataset, we estimate its dip corresponds to a \SWH archival delay: other 2020 public code commits exist, but they had not been archived yet by \SWH at the time.

The exponential growth of the dataset will be relevant when discussing historical trends, like gender ratios, as more recent data points will correspond to exponentially larger amounts of commits.

\subsection{Gender detection}

With these considerations in mind, we proceeded to partition our corpus along the orthogonal gender and world region axes.

To detect the gender of a contribution we use the name of its author, as recorded in the corresponding commit, and apply \PKGGG~\cite{genderguesser} to it.
\PKGGG is an open source Python tool and library for gender detection, which is based on first name frequencies around the world and that is frequently used in related work.
The availability of the package as open source was an important point in favor of its adoption: alternatives based on commercial APIs might give better accuracy (for some world regions), but would both hinder replicability and be very expensive on such a large corpus.
A detailed description and a comparative benchmark of \PKGGG and its main competitors has been conducted by Santamaria and Mihaljevic~\cite{santamaria2018genderapi}.
It shows that \PKGGG works comparatively well with geographically diverse datasets, as ours is by construction.

\PKGGG takes as input a Unicode string, which is expected to be a \emph{first (given) name}, and returns a detected gender among 6 possible values, depending on the tool's confidence about the result: male, mostly male, unknown, mostly female, female, andy (the last one for unisex names).
Author names in our corpus are \emph{full names}, not split into first versus family names.
Aside from the ``API'' mismatch problem, the first/family name distinction is not meaningful across all world cultures represented in our corpus~\cite{ishida2011namesaroundtheworld}.
Hence, to determine the gender of an author we use a \emph{majority criterion}. 
Specifically, we tokenize full name strings into \emph{name tokens}, splitting at each blank, hyphen, or case change (as in CamelCase notation, which we have verified to be used by several authors in the dataset), and then use \PKGGG to determine the gender of each token. 
If and only if a \emph{strict majority} of name tokens for a given author full name is detected as belonging to one gender (no matter how strongly) we associate the majority gender to the author; otherwise their gender will remain unknown.

After this step all commit authors get associated to either a gender or unknown.
As each commit is associated to exactly one author, we can also partition commits by detected gender, by making them inherit the gender of their authors.

\subsection{Region detection}

As shown in \cref{fig:methodology} region detection is performed using two different techniques.
The first geolocation technique, \ALGemail/, uses the country code top-level domain (ccTLD) found in the domain of the email address of commit authors, as recorded by version control system (VCS) commits in our dataset.
We relied on the IANA list of Latin character ccTLDs~\cite{wikipedia-cctld} and manually mapped each corresponding country, sovereign state, or dependent territory to one of the world regions in \cref{fig:worldmap}.

The second geolocation technique, \ALGtzname/, uses the UTC offset and author name of each commit to detect the most likely world region in which the commit was authored.
The UTC offset is the difference in minutes between Coordinated Universal Time (UTC) and local time at a particular place on Earth.
In the initial dataset each commit is associated to a UTC offset.
From it we determine a list of compatible places that, at the time of the commit timestamp, had the same UTC offset as that of the commit.
For making this determination we use the IANA time zone database (or \emph{zoneinfo})~\cite{tzdata}, which includes as places countries and territories worldwide.

For example the following places (with their zoneinfo name in parentheses) had a UTC offset of +240 minutes when their local time was \texttt{2012-01-01 12:00:00}: Russia MSK (Europe/Moscow), Azerbaijan (Asia/Baku), United Arab Emirates (Asia/Dubai), Russia SAMT (Europe/Samara), Oman (Asia/Muscat), Georgia (Asia/Tbilisi), Armenia (Asia/Yerevan), Mauritius (Indian/Mauritius).
When considering local time \texttt{2012-08-01 12:00:00} Azerbaijan is removed from the list because the country moves to a different offset due to daylight saving time (DST).
Then, on local time \texttt{2016-08-01 12:00:00}, Russia MSK is not included anymore, as whole Russia stopped using DST in 2014 and Samara changed offset in 2016.

Then we assign to each compatible place a score that captures the likelihood that a given author name is assigned to a person in that place.
To this end we use a dataset of the frequencies of the most common first and family names by Forebears which, quoting from~\cite{forebear-names}:
\begin{quote} provides the approximate incidence of forenames and surnames produced from a database of \num{4 044 546 938} people (55.5\% of living people in 2014). As of September 2019 it covers \num{27 662 801} forenames and \num{27 206 821} surnames in 236 jurisdictions.\end{quote}
As for gender detection, lacking a first/family name split, we first tokenize names as before and then lookup individual name tokens in both first and family names frequency lists.

For each element found in name lists we multiply the place population\footnotemark{} with the name frequency to obtain a measure that is proportional to the number of persons bearing that name (token) in the specific place.
\footnotetext{To obtain population totals, as the notion of ``place'' at hand is heterogeneous, from a full country to a slice of a larger country spanning multiple timezones, we used a mixture of primary sources (e.g., government websites), and non-primary ones (e.g., Wikipedia articles).}
We sum this figure for all elements to obtain a place score ending up with a list of $\langle$place, score$\rangle$ pairs. 
We then partition this list by the world region that a place belongs to and sum the score for all the places in each region to obtain an overall score, corresponding to the likelihood that the commit belongs to a given world region.
We assign the starting commit as coming from the world region with the highest score.

The two geolocation techniques---\ALGemail/ and \ALGtzname/---suffer from different weaknesses.
The main problem with email is recall: for only about \DATACommitsWithEmail/ of the commits it was possible to associate an author email with a ccTLD.
While at the scale of the full dataset an eviction of about \DATACommitsWithoutEmail/ leaves a very large dataset to work with ($\approx$\,300\,M commits), when projecting to specific years and world regions we have to deal with troublesomely small sample sizes.

As for \ALGtzname/ a relevant issue is that of the UTC offset 0, or time zone zero (TZZ), which is overrepresented in the dataset.
This is due to a number of reasons, including: wrong time zone settings in development environments, incorrect timestamp migrations when a repository was converted across VCS systems (e.g., from Subversion to Git), archival errors, and more.
The over-representation of TZZ commits in the dataset decreases over time: 2000 had \DATACommitsTZZTwoThousand/ TZZ commits, 2010 had \DATACommitsTZZTwoThousandTen/, and 2020 only \DATACommitsTZZTwoThousandTwenty/ (with a dataset grown about 3 orders of magnitude since 2000).

Due to their respective strength and weaknesses, we analyzed commits using both geolocation techniques, as well as a mixture of them (such as using \ALGemail/ for TZZ commits and \ALGtzname/ for the rest) and compared results.
We obtained results that, although with noticeable differences in absolute values, exhibit consistent long-term trends, with the sole exception of cases in which the sample size is very small (10 commits or less).

 \section{Results}
\label{sec:results}

\begin{figure}
  \def\pieheight{3.1cm}
  \centering
  \includegraphics[height=\pieheight]{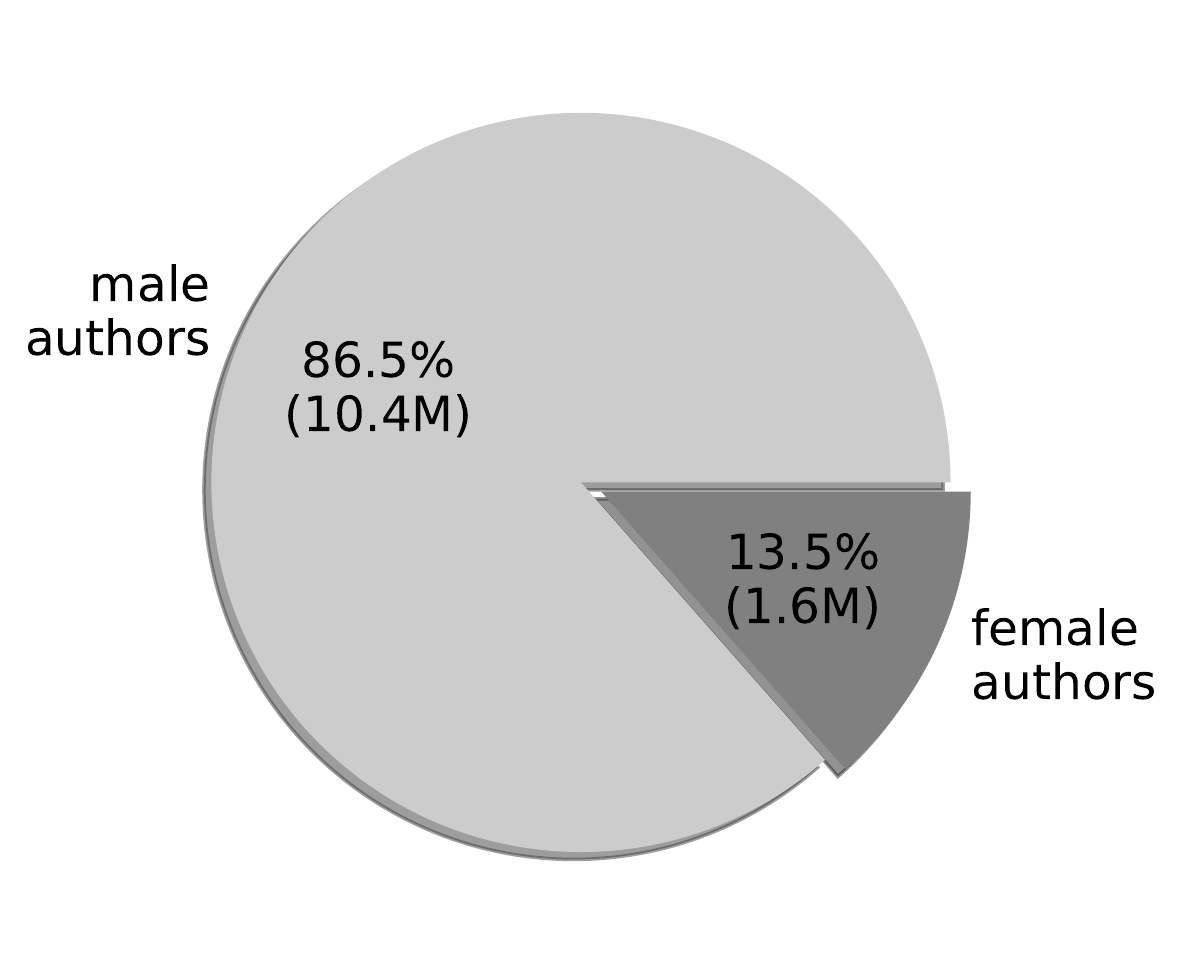}
  \hspace{-3mm}
  \includegraphics[height=\pieheight]{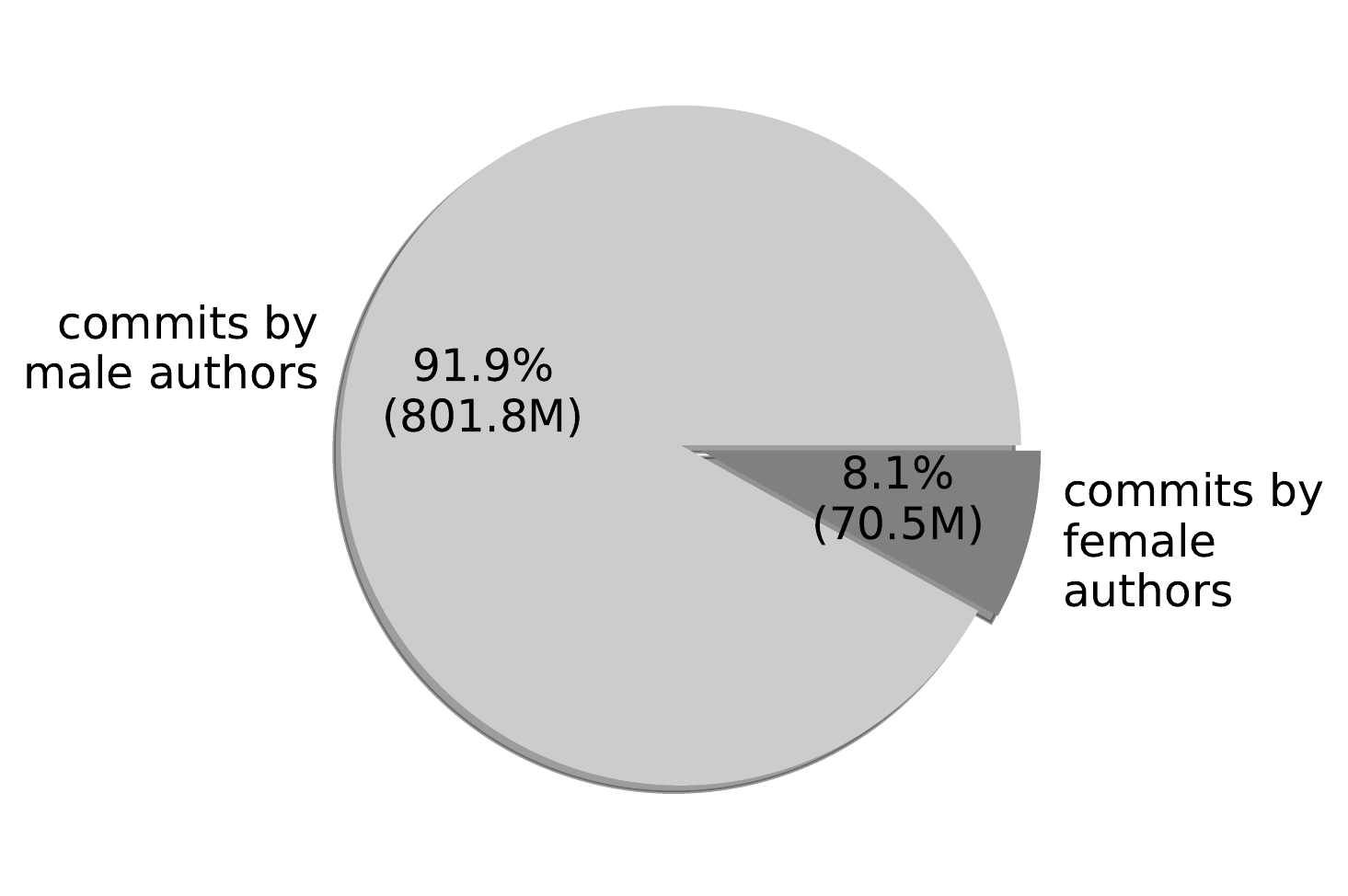}
  \caption{Breakdown of the entire corpus by detected gender for authors (left) and authored commits (right), excluding those for which gender could not be determined.}
  \label{fig:gender-breakdown}
\end{figure}

Using the technique (\PKGGG + majority criterion) discussed in \cref{sec:methodology} we were able to detect the gender for \DATAKnownAuthorsPct/ of the authors in the dataset, corresponding to \DATAKnownAuthorsApprox/ author names.
Those authors have contributed \DATAKnownCommitsPct/ of the commits in the dataset, for a total of \DATAKnownCommitsApprox/, which we can further partition by gender.

\Cref{fig:gender-breakdown} shows the extent of the gender gap in our corpus as a whole, after the exclusion of unknown-gender authors and commits, but  before any other subsampling.
\textbf{The vast majority (\DATAMaleAuthorsPct/) of authors is detected as being male for \DATAMaleAuthorsApprox/ authors v.~only \DATAFemaleAuthorsApprox/ female authors.}
\textbf{The imbalance is even larger when looking at commits, where \DATAMaleCommitsPct/ of commits (for a total of \DATAMaleAuthorsApprox/ commits) have been authored by men v.~\DATAFemaleCommitsPct/ (\DATAFemaleAuthorsApprox/ commits) by women.}

But how does this massive imbalance evolve over time, and how does it change around the world?

\subsection{Gender gap by UTC offset}
\label{sec:gendergapoffset}

We start by answering \cref{rq:gendertimezone}, which is a first approximation of contributor locations on the east-west world axis.

We remind that, by looking only at commit UTC offsets, as needed to answer \cref{rq:gendertimezone}, we lack any information about \emph{timezones}, because they change over time (e.g., due to daylight saving time (DST)) and depend on country-specific regulations.
For what is worth, since DST is mainly adopted in the northern hemisphere from March to November, we repeated the analysis discussed below filtering out all commits performed over that period (which should not introduce relevant gender bias), obtaining results that are analogous to those presented below for the entire dataset.

\begin{figure*}
  \centering
  \hspace*{\fill}
  \subfigure[Commits]{\includegraphics[height=0.95\textheight]{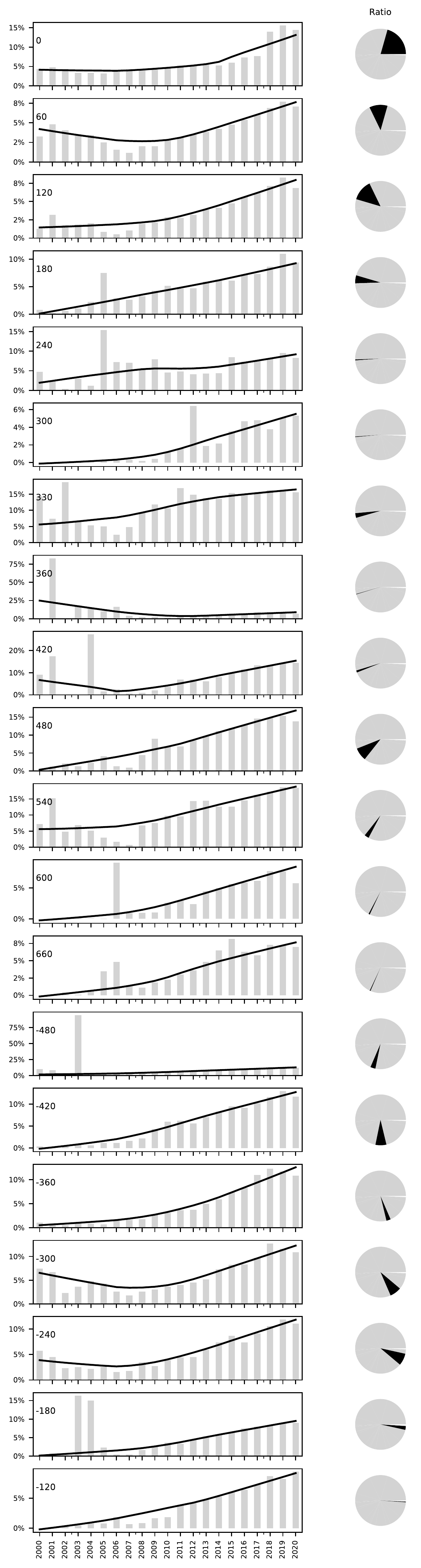} 
    \label{fig:stackedcommitstz}
  }
  \hspace*{\fill}
  \subfigure[Authors]{\includegraphics[height=0.95\textheight]{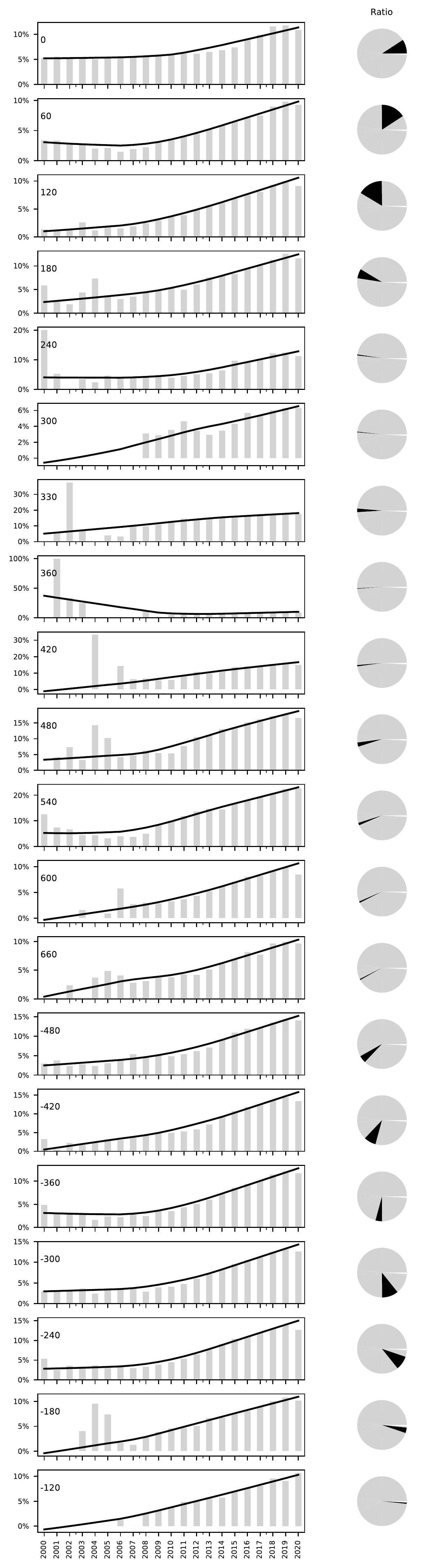}
    \label{fig:stackedpeopletz}
  }
  \hspace*{\fill}
  \caption{Ratio of yearly female authors (right) and their commits (left) for several UTC offsets, starting at 0 and moving east.}
  \label{fig:stackedtz}
\end{figure*}

\Cref{fig:stackedcommitstz} shows the evolution of the ratio of commits by female authors over the \TZANALYZED/ most active time offsets (in terms of available commits in the dataset), displayed as a stacked bar chart.
The solid line in each chart is a loess regression curve~\cite{cleveland1979loess} that materializes the data trend.
The small pie chart to the right of each diagram shows, in black, the ratio of the number of commits that is contributed by that time offset to the entire dataset.

In the following we show and discuss only trends for the period 2000--2020.
Data and charts for the 1970--2000 are available in the replication package, but fall into the pitfall of (1) dealing with partitions derived from an exponentially smaller starting dataset (see \cref{fig:growth}), which (2) is further partitioned by UTC offsets, resulting in slices that are so small to be of limited significance.

\Cref{fig:stackedpeopletz} shows the same evolution, but this time in terms of the number of distinct \emph{authors}, as opposed to the number of authored \emph{commits}.
To avoid outliers due to sporadic contributors we only count authors that have contributed at least 5 commits in a given year.
This threshold is a compromise between the ability to filter out anomalies introduced by drive-by contributors and the significance of the data after filtering.
The value we adopted is the result of a qualitative process started with a large threshold, iteratively lowered until no appreciable outliers were present and yet all data slices were still reasonably represented.
The replication package also includes charts obtained with different threshold values.

We propose these two views---commits in \cref{fig:stackedcommitstz} and authors in \cref{fig:stackedpeopletz}---to overcome inherent potential biases when only discussing one of the two metrics.
Indeed, commit-based gender imbalance can be influenced by prolific contributors (i.e., authors contributing a large number of commits in a given year), whereas author-based gender balance can be influenced by a large number of less prolific contributors.

Note that the stacked graphs for different UTC offsets have different maximum values on their Y axes and hence difference scales.
This is because name-based gender detection is subject to appreciable biases when operating in different world cultures, due to phenomena that include: the adoption of typical female names also for males, the presence of gender-neutral names, a gender-imbalanced reference database, and others.
This makes the comparison of the \emph{absolute value} of ratios between different world zones of little significance, without undermining the comparison of the \emph{evolution trends} of these ratios, because the gender detection approach does not vary over time.
This is what the stacked charts of \cref{fig:stackedtz} allow to do, without having to vertically squeeze some charts too much due to a common scale.

With one exception (discussed below) both views highlight a common trend.
\textbf{The participation of women to the production of public code has grown steadily over the past 12 years, across all UTC offsets, in terms of both contributed commits and active yearly authors.}

The sole exception is offset UTC+240, where the trend is less clear in the authors chart, due to the dip in the 2011--2014 period; after that, the growth is confirmed for 2015--2020.
We note that Russian regions are by far the most populated among those found in this UTC offset and that the years are compatible with when Russia changed their DST adoption, resulting in regions changing UTC offsets.
We therefore speculate that this irregular behavior might be caused by a drastic change of areas (and population) included in this dataset slice.

Offset +360 also shows a recent consistent growth, but the overall commits trend is dominated by the spike in 2001, making subsequent years appear low in comparison.
Looking at the data for 2001 in this offset shows the presence of a single strong female author and only other 6 less prolific authors.
Moreover, the name of the strong author in question, although detected as female with our approach, is a name used for both genders in most of the countries at those latitudes.
A similar effect is visible in the commit chart at offset -480, which is dominated by a 2003 spike.
A detailed analysis of the data shows the large influence of a single strong author (who contributed \num{495785} commits), detected as female but likely either a \emph{bot} or somebody committing changes authored by others.

Overall, the charts also show a growth trend for the 2000--2007 period, although a less steep one and less ubiquitous across UTC offsets.
We observe that, due to exponential dataset growth, the amount of 2008--2020 commits is about 25 times larger than that of 2000--2007, which makes it far easier for the data of less populated regions to be subject to erratic trends due to specific outliers.
For example, the previously discussed commit spike at +240 in 2005 is due to two authors who together contributed 92\% of the commits performed by all women that year.
This is the kind of effect that can affect the commits chart; we expect this not to affect the authors chart and, indeed, \cref{fig:stackedpeopletz} shows no signs of it.

Another outlier is at offset +420 in 2005 (along with two minor ones in 2000 and 2001) that reverses the trends.
A detailed look at the data shows a very small data sample (23 commits in total) and a case of doubtful gender detection.
Outliers aside some offsets (including the well-represented +60) show a 2000--2008 trend of stable and even \emph{decreasing ratio} of women contributions, with no obvious anomalies in the data, which lead to wonder if there are underlying \emph{regional} phenomena apply.

\subsection{Gender gap by world region}
\label{sec:gendergapregion}

\begin{figure*}
  \centering
  \hspace*{\fill}
  \subfigure[Commits]{\includegraphics[height=0.8\textheight]{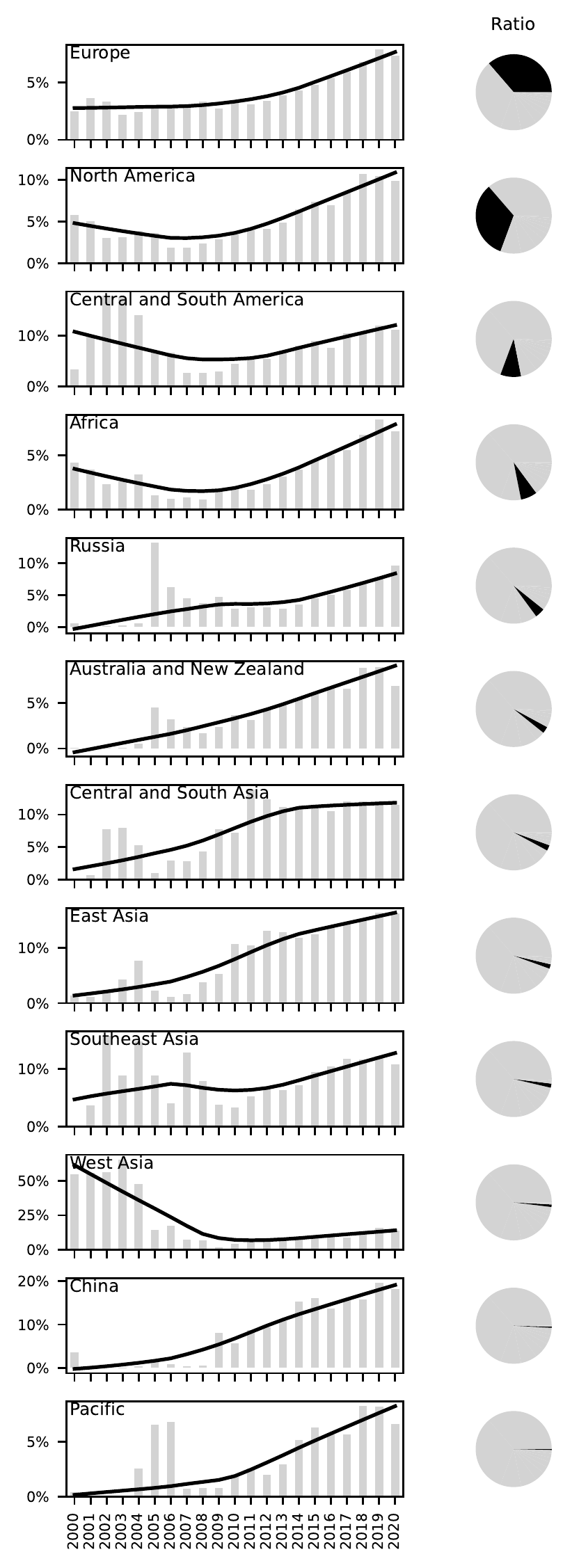} 
    \label{fig:stackedcommitszone}
  }
  \hspace*{\fill}
  \subfigure[Authors]{\includegraphics[height=0.8\textheight]{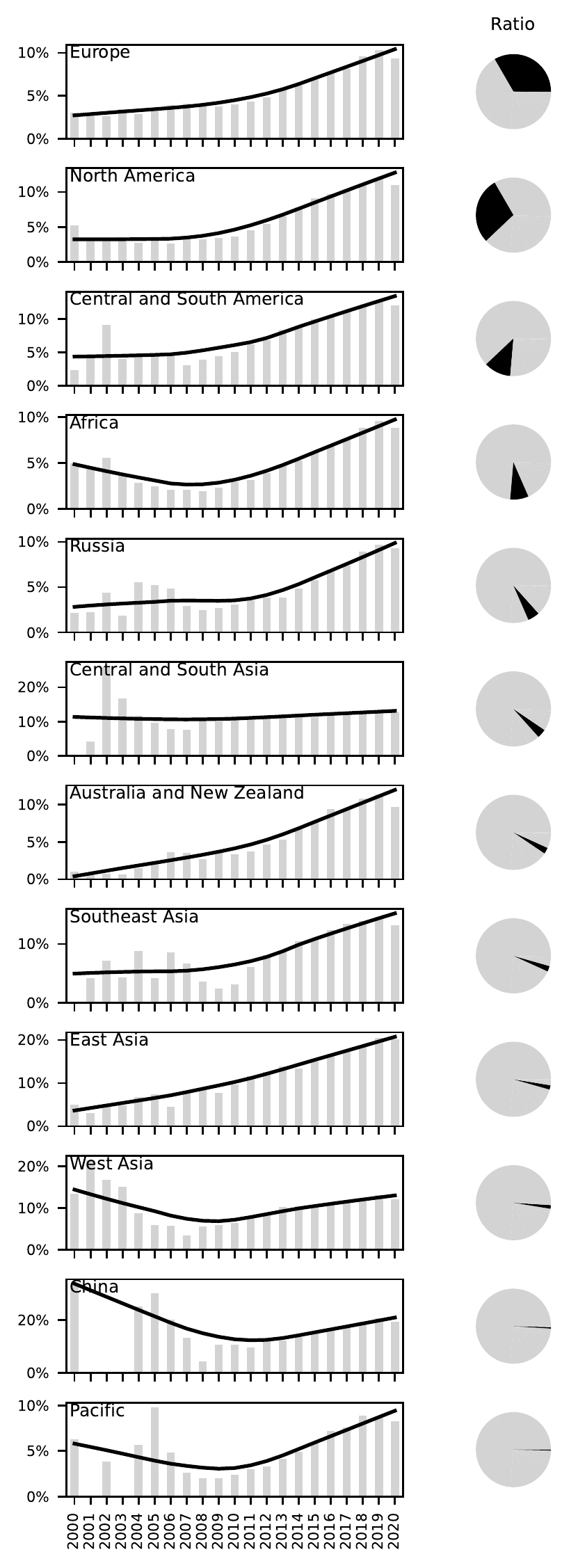}
    \label{fig:stackedpeoplezone}
  }
  \hspace*{\fill}
  \caption{Ratio of yearly female authors (right) and their commits (left) by world region.}
  \label{fig:stackedzone}
\end{figure*}

We explore this aspect by addressing \cref{rq:genderworld}.
For the reasons discussed in \cref{sec:methodology} we use a mixed strategy to assign authors to the world regions of \cref{fig:worldmap}: for commits with UTC offset zero we rely on the \ALGemail/ strategy, whereas for other commits we use the \ALGtzname/ strategy.
\Cref{fig:stackedzone} shows the ratio of both yearly female authors and that of commits authored by women as a set of stacked charts, this time broken down by detected world regions rather than by UTC offset (as it was the case in \cref{fig:stackedtz}).
Previously discussed caveats---about different Y-axis scales and the filtering on active authors by at least 5 yearly commits---still apply to these charts.

The breakdown by world regions confirm the presence of a stable growth trend, with few exceptions (discussed below).
\textbf{Over the past 12 years both the ratio of female authors and that of their commits have grown steadily across most world regions.
Relevant differences across regions, in both the volume and tendencies of the trends, are noticeable} this time.

The recent trend for Central and South Asia (which includes India) shows a very slow growth which even seems to have plateaued in the last few years.
A slow growth (at least in one of the two studied metrics) also characterizes West Asia, recent years in Southeast Asia, and partially China.
Taken together these seem to suggest that Asian zones are subject to a different recent dynamic characterized by a slower growth of female contribution.

When taking into account the whole 2000--2020 time span, it appears that the growth that characterizes 2008--2020 was not present in 2000--2007.
In some cases growth is less evident, in some cases there is even a reduction in the ratio of women participation, resulting in an overall trend with an upward concave curve hitting a minimum at about 2008.

Further investigation shows that for some of the regions this behavior can originate from local anomalies.
For instance, the West Asia subsample is dominated by Israel, where \PKGGG detects as female \emph{``Eli''}, which is an Hebraic male name that is also incidentally the name for several prolific committers in those years. 
China is subject to effects related to being a very limited subsample: the only region with less commits than China is Pacific. 
This translates, for the first years of the studied period, to a very limited number of commits and authors (less than a dozen is some cases) so the overall trend can be severely affected by even small anomalies.
In all other cases there seem to be enough good-quality data to support the conclusion of a legitimate shrinking/expanding trend, which can be seem for commits (and partially authors) in the Americas, South East Asia, and Africa.

Overall, we find evidence that female participation to public code production has been growing stably in most world regions over the past 12 years, with a less pronounced trend in some Asian countries.
Looking further back to the past 20 years, women participation to public code is subject to different regional developments.

\subsection{Gender gap and the COVID-19 pandemic}
\label{sec:gendergapcovid}

\begin{figure}
  \centering
  \includegraphics[height=0.8\textheight]{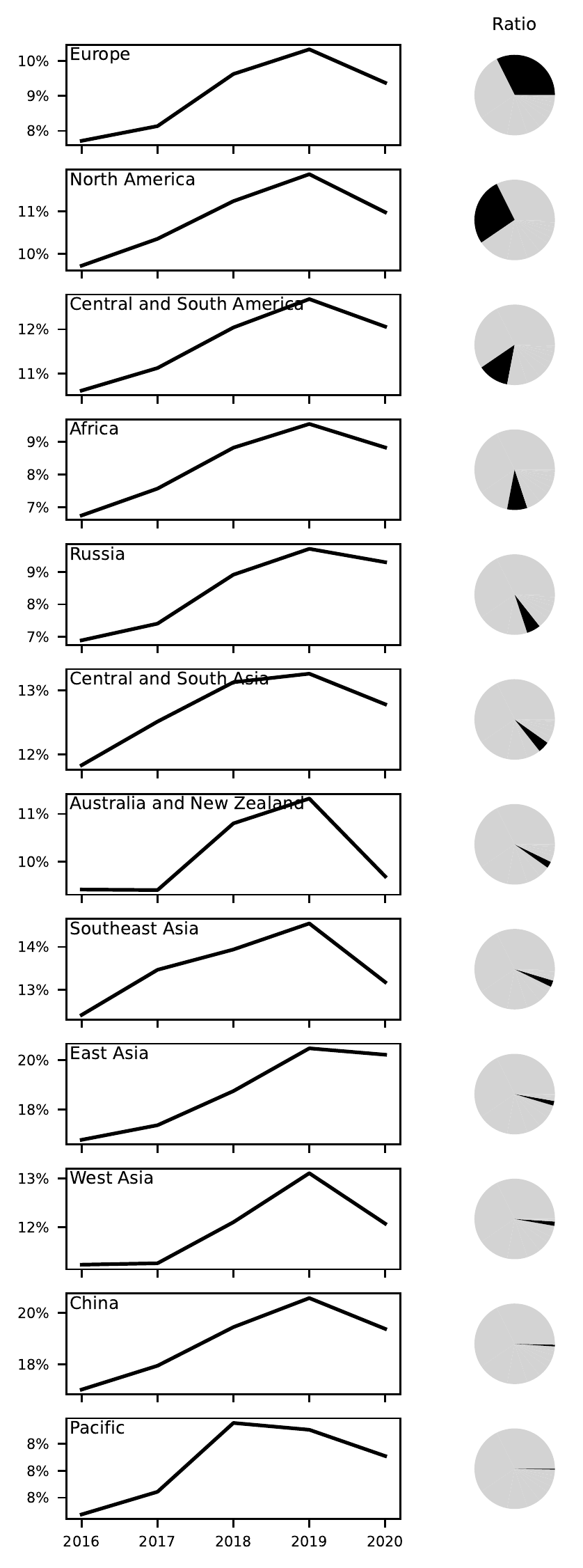}
  \caption{Ratio of yearly female authors during the 2016--2020 period, by world region.
    Women participation has decreased everywhere in 2020 w.r.t.~2019.
  }
  \label{fig:stackedpeople1620}
\end{figure}

Previous results consistently show a marked growth trend in female participation to the production of public code over the last 12 years. 
However, when looking closely at previous charts a recurrent worldwide anomaly also jumps to the eye: \textbf{women participation in the production of public code has decreased everywhere in 2020 with respect to 2019}.
Whereas this decrease is not enough to change the loess trend, it is nevertheless present.
More precisely: across all world regions the ratio of yearly active female authors has decreased in 2020 w.r.t.~2019; and across all regions but one, the ratio of commits contributed by women has decreased in 2020 w.r.t.~2019.
To better highlight this phenomenon we visualize it in \cref{fig:stackedpeople1620}, for authors only, zooming into the 2016--2020 period, and using stacked line charts.
Note that this decrease is in stark contrast with the worldwide trend of \emph{increased} women participation over the past 12 years.
Remember also that the observed variations are in \emph{ratios}, which could not be explained by an incomplete dataset or other cross-cutting phenomena: after getting better for almost 12 years the gender gap has worsened again in 2020. 

While not originally part of our study design, this recurrent anomaly led us to state \cref{rq:gendercovid} about the impact of the COVID-19 pandemic on women participation in the production of public code.
The decrease in women participation in 2020 stands out as an anomaly in the observed trends up to 2019 and is \emph{correlated} with the insurgence of the pandemic in 2020.
While with the data we have at hand we cannot verify a causal relation between the two, it is our educated guess that this decrease has been caused by the COVID-19 pandemics.
The closures of schools and daycare facilities have increased responsibilities for caregivers which, as many studies have shown (among them~\cite{alon2020covidgender, collins2021covidgender}), have not been equally split between genders, impacting more on women's ability to continue working than men's.
A plausible interpretation of our data is hence that women's ability to contribute to public code---either as part of their day job, or as an activity conducted during their spare time---has also been negatively impacted by the COVID-19 pandemic, increasing the gender gap in 2020.

Interestingly enough, the impact of this imbalance seems to be more limited in Asiatic regions than elsewhere in the world. 
Whether this is due to a different impact of the pandemic on gender imbalance in those societies, or to different local contribution patterns to public code between genders, is something we cannot assess and that remains to be explored as future work.

 \section{Limitations}
\label{sec:limitations}
\label{sec:threats}

In this section we discuss threats to the validity of the obtained results,
starting with construct validity concerns.

\subsection{Construct validity}

\paragraph{Accuracy/scale trade-off}
This is the first study, to the best of our knowledge, that has analyzed at this scale the geographic dimension of the gender gap in public code contributions.
Dealing with such a large dataset (\AuthorCount authors and \CommitCount commits, before any filtering) calls for the crude, but fully automated methods, that we have adopted for detecting both author genders and their coarse-grained location.
Alternative approaches found in the literature rely on crawling individual author information (e.g., from social media or development platform profiles) and even one-to-one interviews with them.
There appears to be a clear trade-off between the accuracy of the latter methods and the ability to scale to much larger datasets of the methods we have chosen for in this study.

Considering that: (1) we have relied for the most parts on methods and techniques that are also found in the literature (even though they have not been applied at this scale before), (2) we are only drawing conclusions about long-term aggregate trends, and (3) the trends we observe appear to be statistically stable; we do not consider our general choice of analysis methods a significant threat to the validity of our answers to the stated research questions.
It is still worth reviewing specific methodological choices.

\paragraph{Gender detection}
Regarding gender detection, the choice of\linebreak \PKGGG as building block is based on a preexisting benchmark~\cite{santamaria2018genderapi} of automated gender detection tools and on the fact that, being open source, it enables replicating our findings without depending on third-party APIs.
On top of the tool itself we added a majority criterion, due to the need of working with non-parsed author names.
It is trivial to come up with handcrafted cases that break this heuristic, for example with \emph{family} names composed by name tokens that are also common \emph{first} names detected as belonging to the ``wrong'' gender w.r.t.~actual author gender (which is a phenomenon mostly affecting Chinese names).
Other than that, though, most family names are detected by \PKGGG as being of an ``unknown'' gender, which does not skew the gender majority of an author in any direction.
Hence in practice, especially at this scale, we do not expect this aspect to significantly impact experimental results.
Another limit of \PKGGG only marginally impacts our study: it only operates on latin alphabet names making it unapplicable to names adopting other alphabets.
However we observed that most developers from countries such as Japan, China, South Korea, Thailand and others, usually adopt Western names when contributing to public code, limiting the impact of the problem.

The proposed heuristic is unable to determine a gender for a large part of the starting dataset (\DATAUnknownAuthorsPct/ authors and \DATAUnknownCommitsPct/ commits), which might impact results.
The remaining part of the dataset still corresponds to the largest scale study of this kind, which we believe is important to report about.
We have also mitigated this risk by just looking at trend ratios within the subset of authors for which we could determine a gender.
It will be trivial in the future to integrate upcoming improvements in name-based gender detection in replications of our experiments.

\paragraph{Region detection}

The two techniques used for region detection has complementary strengths and weaknesses.
On the one hand, ccTLD-based region detection, in addition to having limited applicability, is affected by the Internet practices that surround the use of national top-level domains (TLDs).
Some of them are rarely used, such as \texttt{.us} for the USA, where generic TLDs (\texttt{.com}, \texttt{.net}, \texttt{.org}, etc.) originated from and remain to this date much more popular than elsewhere.
As a consequence ccTLD-based region detection leads to the underrepresentation of countries like the USA.
Conversely some ccTLDs are used worldwide as part of popular ``domain hacks'', e.g., \texttt{.io} from British Indian Ocean Territory used for computer-related websites, or \texttt{.tv} from Tuvalu used for television ones.

On the other hand, \ALGtzname/-based region detection relies on population totals retrieved from different and potentially heterogeneous public sources.
In particular official population figures are usually aligned with national census that, in some cases, can be up to 15 years old.
As a consequence we might have ended up comparing, say, 2020 population figures with 2005 figures.
For places with relatively stable population this is not an issue, but it could become one in more unstable regions.

We expect the impact of these weaknesses to be marginal in our dataset overall.
Nonetheless we have mitigated their potential effects by combining the two techniques as detailed in \cref{sec:methodology}.
We have also compared results obtained with each technique \emph{separately}, obtaining similar results outside of the peculiar timezone zero.

\subsection{External validity}

The dataset we have analyzed does not correspond to the fully body of neither public code nor free/open source software.
Our findings hence inherit the limitations of \SWH as a research archive, in particular related to the (lack of) coverage of specific development platforms or software distribution technologies, and to archival lag (which we have observed for the last ``complete'' year in our dataset: 2020).
The study dataset is, however, the largest approximation of contributions to public code that is readily and publicly available for analysis, which we believe validates our choice of starting point.
Larger samples of public code can be analyzed (and probably will in the future), but we do not expect significant differences to emerge from similar incremental improvements in coverage.
It would be more relevant to complement our analysis adding significant bodies of \emph{non-public code}, such as those developed via large in-house private forges, which can potentially constitute a very different population in term of gender gap evolution.
We make no claim about the generalizability of our findings to those contexts.

 \section{Conclusion}
\label{sec:conclusion}

We have studied the gender gap in public code contributions along the orthogonal axes of time and geographic location of contributors.
To that end, we have used heuristics based on name frequencies, email domains, and timestamp offsets, that enabled us to analyze \CommitCount commits contributed by \AuthorCount authors over a period of 50 years.
We confirm previous results about the gender gap in public code: women have contributed less than 10\% of public code overall, but the ratio of their involvement is growing steadily.
We provide novel evidence that this growth---both in terms of active female authors and of their commits---is a global trend, shared by most world regions over the past 12 years.
However, 2020 has been a setback year, with the ratio of women participation decreasing everywhere in the world, most likely due to the COVID-19 pandemic disproportionately affecting women.

\paragraph{Future work}

As future work, region-specific analyses to understand local trends would be useful in particular, but not only, to understand why the COVID-19 contraction in women participation appears to have impacted Asiatic regions less than others.
The ability to study large bodies of non-public, but still collaboratively developed, code is also to be pursued, in order to compare collaboration dynamics before versus away from the public eye.

We also intend to attempt large-scale validation, which remains challenging on datasets of this scale, as well as quantitative comparisons with findings by other studies in selected world region.
Provided that gender diversity results for \emph{identifiable subsets} of the population analyzed in this study are available from other studies, one can cross-check results to either reinforce the respective findings or pinpoint the causes of discrepancies, informing future studies.


\clearpage


\begin{thebibliography}{42}

\ifx \showCODEN    \undefined \def \showCODEN     #1{\unskip}     \fi
\ifx \showDOI      \undefined \def \showDOI       #1{#1}\fi
\ifx \showISBNx    \undefined \def \showISBNx     #1{\unskip}     \fi
\ifx \showISBNxiii \undefined \def \showISBNxiii  #1{\unskip}     \fi
\ifx \showISSN     \undefined \def \showISSN      #1{\unskip}     \fi
\ifx \showLCCN     \undefined \def \showLCCN      #1{\unskip}     \fi
\ifx \shownote     \undefined \def \shownote      #1{#1}          \fi
\ifx \showarticletitle \undefined \def \showarticletitle #1{#1}   \fi
\ifx \showURL      \undefined \def \showURL       {\relax}        \fi
\providecommand\bibfield[2]{#2}
\providecommand\bibinfo[2]{#2}
\providecommand\natexlab[1]{#1}
\providecommand\showeprint[2][]{arXiv:#2}

\bibitem[\protect\citeauthoryear{Alon, Doepke, Olmstead-Rumsey, and
  Tertilt}{Alon et~al\mbox{.}}{2020}]%
        {alon2020covidgender}
\bibfield{author}{\bibinfo{person}{Titan Alon}, \bibinfo{person}{Matthias
  Doepke}, \bibinfo{person}{Jane Olmstead-Rumsey}, {and}
  \bibinfo{person}{Mich{\`e}le Tertilt}.} \bibinfo{year}{2020}\natexlab{}.
\newblock \bibinfo{booktitle}{\emph{The impact of {COVID-19} on gender
  equality}}.
\newblock \bibinfo{type}{{T}echnical {R}eport}. \bibinfo{institution}{National
  Bureau of Economic Research}.
\newblock
\newblock
\shownote{Working paper available at:
  \url{https://www.nber.org/system/files/working_papers/w26947/w26947.pdf},
  accessed on 2021-10-13}.


\bibitem[\protect\citeauthoryear{Bosu and Sultana}{Bosu and Sultana}{2019}]%
        {bosu2019fossdiversity}
\bibfield{author}{\bibinfo{person}{Amiangshu Bosu} {and}
  \bibinfo{person}{Kazi~Zakia Sultana}.} \bibinfo{year}{2019}\natexlab{}.
\newblock \showarticletitle{Diversity and Inclusion in Open Source Software
  {(OSS)} Projects: Where Do We Stand?}. In \bibinfo{booktitle}{\emph{2019
  {ACM/IEEE} International Symposium on Empirical Software Engineering and
  Measurement, {ESEM} 2019, Porto de Galinhas, Recife, Brazil, September 19-20,
  2019}}. \bibinfo{publisher}{{IEEE}}, \bibinfo{pages}{1--11}.
\newblock
\urldef\tempurl%
\url{https://doi.org/10.1109/ESEM.2019.8870179}
\showDOI{\tempurl}


\bibitem[\protect\citeauthoryear{Botella, Rueda, L{\'o}pez-I{\~n}esta, and
  Marzal}{Botella et~al\mbox{.}}{2019}]%
        {botella2019genderstem}
\bibfield{author}{\bibinfo{person}{Carmen Botella}, \bibinfo{person}{Silvia
  Rueda}, \bibinfo{person}{Emilia L{\'o}pez-I{\~n}esta}, {and}
  \bibinfo{person}{Paula Marzal}.} \bibinfo{year}{2019}\natexlab{}.
\newblock \showarticletitle{Gender diversity in {STEM} disciplines: A multiple
  factor problem}.
\newblock \bibinfo{journal}{\emph{Entropy}} \bibinfo{volume}{21},
  \bibinfo{number}{1} (\bibinfo{year}{2019}), \bibinfo{pages}{30}.
\newblock


\bibitem[\protect\citeauthoryear{Canedo, Bonif{\'{a}}cio, Okimoto, Serebrenik,
  Pinto, and Monteiro}{Canedo et~al\mbox{.}}{2020}]%
        {canedo2020womencoredev}
\bibfield{author}{\bibinfo{person}{Edna~Dias Canedo}, \bibinfo{person}{Rodrigo
  Bonif{\'{a}}cio}, \bibinfo{person}{M{\'{a}}rcio~Vinicius Okimoto},
  \bibinfo{person}{Alexander Serebrenik}, \bibinfo{person}{Gustavo Pinto},
  {and} \bibinfo{person}{Eduardo Monteiro}.} \bibinfo{year}{2020}\natexlab{}.
\newblock \showarticletitle{Work Practices and Perceptions from Women Core
  Developers in {OSS} Communities}. In \bibinfo{booktitle}{\emph{{ESEM} '20:
  {ACM} / {IEEE} International Symposium on Empirical Software Engineering and
  Measurement, Bari, Italy, October 5-7, 2020}},
  \bibfield{editor}{\bibinfo{person}{Maria~Teresa Baldassarre},
  \bibinfo{person}{Filippo Lanubile}, \bibinfo{person}{Marcos Kalinowski},
  {and} \bibinfo{person}{Federica Sarro}} (Eds.). \bibinfo{publisher}{{ACM}},
  \bibinfo{pages}{26:1--26:11}.
\newblock
\urldef\tempurl%
\url{https://doi.org/10.1145/3382494.3410682}
\showDOI{\tempurl}


\bibitem[\protect\citeauthoryear{Chavatzia}{Chavatzia}{2017}]%
        {unesco2017genderstem}
\bibfield{author}{\bibinfo{person}{Theophania Chavatzia}.}
  \bibinfo{year}{2017}\natexlab{}.
\newblock \bibinfo{booktitle}{\emph{Cracking the code: Girls' and women's
  education in science, technology, engineering and mathematics ({STEM})}}.
\newblock \bibinfo{type}{{T}echnical {R}eport}.
  \bibinfo{institution}{{UNESCO}}.
\newblock
\urldef\tempurl%
\url{https://unesdoc.unesco.org/ark:/48223/pf0000260079}
\showURL{%
\tempurl}
\newblock
\shownote{Online; accessed 2022-02-08}.


\bibitem[\protect\citeauthoryear{Cleveland}{Cleveland}{1979}]%
        {cleveland1979loess}
\bibfield{author}{\bibinfo{person}{William~S. Cleveland}.}
  \bibinfo{year}{1979}\natexlab{}.
\newblock \showarticletitle{Robust Locally Weighted Regression and Smoothing
  Scatterplots}.
\newblock \bibinfo{journal}{\emph{J. Am. Stat. Assoc.}} \bibinfo{volume}{74},
  \bibinfo{number}{368} (\bibinfo{date}{Dec} \bibinfo{year}{1979}),
  \bibinfo{pages}{829--836}.
\newblock
\showISSN{0162-1459}
\urldef\tempurl%
\url{https://doi.org/10.1080/01621459.1979.10481038}
\showDOI{\tempurl}


\bibitem[\protect\citeauthoryear{Collins, Landivar, Ruppanner, and
  Scarborough}{Collins et~al\mbox{.}}{2021}]%
        {collins2021covidgender}
\bibfield{author}{\bibinfo{person}{Caitlyn Collins},
  \bibinfo{person}{Liana~Christin Landivar}, \bibinfo{person}{Leah Ruppanner},
  {and} \bibinfo{person}{William~J. Scarborough}.}
  \bibinfo{year}{2021}\natexlab{}.
\newblock \showarticletitle{{COVID-19} and the gender gap in work hours}.
\newblock \bibinfo{journal}{\emph{Gender, Work \& Organization}}
  \bibinfo{volume}{28}, \bibinfo{number}{S1} (\bibinfo{year}{2021}),
  \bibinfo{pages}{101--112}.
\newblock
\urldef\tempurl%
\url{https://doi.org/10.1111/gwao.12506}
\showDOI{\tempurl}


\bibitem[\protect\citeauthoryear{David and Shapiro}{David and Shapiro}{2008}]%
        {david2008fossdevs}
\bibfield{author}{\bibinfo{person}{Paul~A David} {and}
  \bibinfo{person}{Joseph~S Shapiro}.} \bibinfo{year}{2008}\natexlab{}.
\newblock \showarticletitle{Community-based production of open-source software:
  What do we know about the developers who participate?}
\newblock \bibinfo{journal}{\emph{Information Economics and Policy}}
  \bibinfo{volume}{20}, \bibinfo{number}{4} (\bibinfo{year}{2008}),
  \bibinfo{pages}{364--398}.
\newblock


\bibitem[\protect\citeauthoryear{Di~Cosmo and Zacchiroli}{Di~Cosmo and
  Zacchiroli}{2017}]%
        {swhipres2017}
\bibfield{author}{\bibinfo{person}{Roberto Di~Cosmo} {and}
  \bibinfo{person}{Stefano Zacchiroli}.} \bibinfo{year}{2017}\natexlab{}.
\newblock \showarticletitle{{Software Heritage}: Why and How to Preserve
  Software Source Code}. In \bibinfo{booktitle}{\emph{Proceedings of the 14th
  International Conference on Digital Preservation, iPRES 2017}}.
\newblock
\urldef\tempurl%
\url{https://hal.archives-ouvertes.fr/hal-01590958/}
\showURL{%
\tempurl}


\bibitem[\protect\citeauthoryear{Ferhat~Elmas}{Ferhat~Elmas}{2015}]%
        {genderguesser}
\bibfield{author}{\bibinfo{person}{Israel Saeta~Pérez Ferhat~Elmas,
  David~Arcos}.} \bibinfo{year}{2015}\natexlab{}.
\newblock \bibinfo{title}{gender-guesser}.
\newblock
  \bibinfo{howpublished}{\url{https://github.com/lead-ratings/gender-guesser}}.
\newblock
\newblock
\shownote{Retrieved 2021-09-28}.


\bibitem[\protect\citeauthoryear{Forebears}{Forebears}{2021}]%
        {forebear-names}
\bibfield{author}{\bibinfo{person}{Forebears}.}
  \bibinfo{year}{2021}\natexlab{}.
\newblock \bibinfo{title}{World Forename {\&} Surname Distribution Maps}.
\newblock
\newblock
\newblock
\shownote{Online at
  \url{https://forebears.io/about/name-distribution-and-demographics}, accessed
  during April 2021}.


\bibitem[\protect\citeauthoryear{Frenkel}{Frenkel}{1990}]%
        {frenkel1990womencs}
\bibfield{author}{\bibinfo{person}{Karen~A Frenkel}.}
  \bibinfo{year}{1990}\natexlab{}.
\newblock \showarticletitle{Women and computing}.
\newblock \bibinfo{journal}{\emph{Commun. ACM}} \bibinfo{volume}{33},
  \bibinfo{number}{11} (\bibinfo{year}{1990}), \bibinfo{pages}{34--46}.
\newblock


\bibitem[\protect\citeauthoryear{Gonz{\'{a}}lez{-}Barahona, Robles,
  Andradas{-}Izquierdo, and Ghosh}{Gonz{\'{a}}lez{-}Barahona
  et~al\mbox{.}}{2008}]%
        {barahona2008geography}
\bibfield{author}{\bibinfo{person}{Jes{\'{u}}s~M. Gonz{\'{a}}lez{-}Barahona},
  \bibinfo{person}{Gregorio Robles}, \bibinfo{person}{Roberto
  Andradas{-}Izquierdo}, {and} \bibinfo{person}{Rishab~Aiyer Ghosh}.}
  \bibinfo{year}{2008}\natexlab{}.
\newblock \showarticletitle{Geographic origin of libre software developers}.
\newblock \bibinfo{journal}{\emph{Inf. Econ. Policy}} \bibinfo{volume}{20},
  \bibinfo{number}{4} (\bibinfo{year}{2008}), \bibinfo{pages}{356--363}.
\newblock
\urldef\tempurl%
\url{https://doi.org/10.1016/j.infoecopol.2008.07.001}
\showDOI{\tempurl}


\bibitem[\protect\citeauthoryear{Gonz{\'{a}}lez{-}Barahona, Robles, and
  Izquierdo{-}Cortazar}{Gonz{\'{a}}lez{-}Barahona et~al\mbox{.}}{2016}]%
        {barahona2016geodistrib}
\bibfield{author}{\bibinfo{person}{Jes{\'{u}}s~M. Gonz{\'{a}}lez{-}Barahona},
  \bibinfo{person}{Gregorio Robles}, {and} \bibinfo{person}{Daniel
  Izquierdo{-}Cortazar}.} \bibinfo{year}{2016}\natexlab{}.
\newblock \showarticletitle{Determining the Geographical distribution of a
  Community by means of a Time-zone Analysis}. In
  \bibinfo{booktitle}{\emph{Proceedings of the 12th International Symposium on
  Open Collaboration, OpenSym 2016, Berlin, Germany, August 17-19, 2016}},
  \bibfield{editor}{\bibinfo{person}{Anthony~I. Wasserman}} (Ed.).
  \bibinfo{publisher}{{ACM}}, \bibinfo{pages}{3:1--3:4}.
\newblock
\urldef\tempurl%
\url{https://doi.org/10.1145/2957792.2957802}
\showDOI{\tempurl}


\bibitem[\protect\citeauthoryear{Hill, Corbett, and St~Rose}{Hill
  et~al\mbox{.}}{2010}]%
        {hill2010whysofew}
\bibfield{author}{\bibinfo{person}{Catherine Hill},
  \bibinfo{person}{Christianne Corbett}, {and} \bibinfo{person}{Andresse
  St~Rose}.} \bibinfo{year}{2010}\natexlab{}.
\newblock \bibinfo{booktitle}{\emph{Why so few? Women in science, technology,
  engineering, and mathematics.}}
\newblock \bibinfo{publisher}{ERIC}.
\newblock


\bibitem[\protect\citeauthoryear{IANA}{IANA}{2017}]%
        {tzdata}
\bibfield{author}{\bibinfo{person}{IANA}.} \bibinfo{year}{2017}\natexlab{}.
\newblock \bibinfo{title}{Time Zone Database}.
\newblock
\newblock
\urldef\tempurl%
\url{https://data.iana.org/time-zones/releases/}
\showURL{%
\tempurl}
\newblock
\shownote{Retrieved 2021-09-28}.


\bibitem[\protect\citeauthoryear{{IANA}}{{IANA}}{2021}]%
        {wikipedia-cctld}
\bibfield{author}{\bibinfo{person}{{IANA}}.} \bibinfo{year}{2021}\natexlab{}.
\newblock \bibinfo{title}{Country code top-level domains}.
\newblock
\newblock
\newblock
\shownote{Mirrored at
  \url{https://en.wikipedia.org/wiki/Country_code_top-level_domain\#Latin_Character_ccTLDs},
  accessed 2021-10-06}.


\bibitem[\protect\citeauthoryear{Ishida}{Ishida}{2011}]%
        {ishida2011namesaroundtheworld}
\bibfield{author}{\bibinfo{person}{Richard Ishida}.}
  \bibinfo{year}{2011}\natexlab{}.
\newblock \bibinfo{title}{Personal names around the world}.
\newblock
  \bibinfo{howpublished}{\url{https://www.w3.org/International/questions/qa-personal-names}}.
\newblock


\bibitem[\protect\citeauthoryear{Izquierdo, Huesman, Serebrenik, and
  Robles}{Izquierdo et~al\mbox{.}}{2019}]%
        {izquierdo2019openstackdiversity}
\bibfield{author}{\bibinfo{person}{Daniel Izquierdo}, \bibinfo{person}{Nicole
  Huesman}, \bibinfo{person}{Alexander Serebrenik}, {and}
  \bibinfo{person}{Gregorio Robles}.} \bibinfo{year}{2019}\natexlab{}.
\newblock \showarticletitle{{OpenStack} Gender Diversity Report}.
\newblock \bibinfo{journal}{\emph{{IEEE} Softw.}} \bibinfo{volume}{36},
  \bibinfo{number}{1} (\bibinfo{year}{2019}), \bibinfo{pages}{28--33}.
\newblock
\urldef\tempurl%
\url{https://doi.org/10.1109/MS.2018.2874322}
\showDOI{\tempurl}


\bibitem[\protect\citeauthoryear{Kuechler, Gilbertson, and Jensen}{Kuechler
  et~al\mbox{.}}{2012}]%
        {kuechler2012genderfoss}
\bibfield{author}{\bibinfo{person}{Victor Kuechler}, \bibinfo{person}{Claire
  Gilbertson}, {and} \bibinfo{person}{Carlos Jensen}.}
  \bibinfo{year}{2012}\natexlab{}.
\newblock \showarticletitle{Gender Differences in Early Free and Open Source
  Software Joining Process}. In \bibinfo{booktitle}{\emph{8th International
  Conference on Open Source Systems, {OSS} 2012}}
  \emph{(\bibinfo{series}{{IFIP} Advances in Information and Communication
  Technology}, Vol.~\bibinfo{volume}{378})}. \bibinfo{publisher}{Springer},
  \bibinfo{pages}{78--93}.
\newblock
\urldef\tempurl%
\url{https://doi.org/10.1007/978-3-642-33442-9\_6}
\showDOI{\tempurl}


\bibitem[\protect\citeauthoryear{Margolis and Fisher}{Margolis and
  Fisher}{2002}]%
        {margolis2002womencs}
\bibfield{author}{\bibinfo{person}{Jane Margolis} {and} \bibinfo{person}{Allan
  Fisher}.} \bibinfo{year}{2002}\natexlab{}.
\newblock \bibinfo{booktitle}{\emph{Unlocking the clubhouse: Women in
  computing}}.
\newblock \bibinfo{publisher}{MIT press}.
\newblock


\bibitem[\protect\citeauthoryear{Nafus}{Nafus}{2012}]%
        {nafus2012patches}
\bibfield{author}{\bibinfo{person}{Dawn Nafus}.}
  \bibinfo{year}{2012}\natexlab{}.
\newblock \showarticletitle{‘Patches don't have gender’: What is not open
  in open source software}.
\newblock \bibinfo{journal}{\emph{New Media \& Society}} \bibinfo{volume}{14},
  \bibinfo{number}{4} (\bibinfo{year}{2012}), \bibinfo{pages}{669--683}.
\newblock


\bibitem[\protect\citeauthoryear{Nations}{Nations}{1999}]%
        {un1999geoscheme}
\bibfield{author}{\bibinfo{person}{United Nations}.}
  \bibinfo{year}{1999}\natexlab{}.
\newblock \bibinfo{booktitle}{\emph{Standard country or area codes for
  statistical use}}.
\newblock \bibinfo{type}{{T}echnical {R}eport}. \bibinfo{institution}{United
  Nations}.
\newblock
\urldef\tempurl%
\url{https://unstats.un.org/unsd/methodology/m49/}
\showURL{%
\tempurl}
\newblock
\shownote{Retrieved 2021-09-27}.


\bibitem[\protect\citeauthoryear{of~Economic and Social~Affairs}{of~Economic
  and Social~Affairs}{2019}]%
        {un2019worldpop}
\bibfield{author}{\bibinfo{person}{Department of Economic} {and}
  \bibinfo{person}{Population~Division Social~Affairs}.}
  \bibinfo{year}{2019}\natexlab{}.
\newblock \bibinfo{booktitle}{\emph{World Population Prospects 2019}}.
\newblock \bibinfo{type}{{T}echnical {R}eport}. \bibinfo{institution}{United
  Nations}.
\newblock
\urldef\tempurl%
\url{https://population.un.org/wpp/Download/Standard/Population/}
\showURL{%
\tempurl}
\newblock
\shownote{Retrieved 2021-09-27}.


\bibitem[\protect\citeauthoryear{O'Neil, Raissi, de~Blanc, and
  Zacchiroli}{O'Neil et~al\mbox{.}}{2017}]%
        {oneil2016debiansurvey}
\bibfield{author}{\bibinfo{person}{Mathieu O'Neil}, \bibinfo{person}{Mahin
  Raissi}, \bibinfo{person}{Molly de Blanc}, {and} \bibinfo{person}{Stefano
  Zacchiroli}.} \bibinfo{year}{2017}\natexlab{}.
\newblock \showarticletitle{Preliminary Report on the Influence of Capital in
  an Ethical-Modular Project: Quantitative data from the 2016 Debian Survey}.
\newblock \bibinfo{journal}{\emph{Journal of Peer Production}}
  \bibinfo{number}{10} (\bibinfo{year}{2017}).
\newblock
\showISSN{2213-5316}


\bibitem[\protect\citeauthoryear{Pietri, Spinellis, and Zacchiroli}{Pietri
  et~al\mbox{.}}{2019}]%
        {swh-msr2019-dataset}
\bibfield{author}{\bibinfo{person}{Antoine Pietri}, \bibinfo{person}{Diomidis
  Spinellis}, {and} \bibinfo{person}{Stefano Zacchiroli}.}
  \bibinfo{year}{2019}\natexlab{}.
\newblock \showarticletitle{The {S}oftware {H}eritage graph dataset: public
  software development under one roof}. In \bibinfo{booktitle}{\emph{16th
  International Conference on Mining Software Repositories, {MSR} 2019}}.
  \bibinfo{pages}{138--142}.
\newblock
\urldef\tempurl%
\url{https://dl.acm.org/citation.cfm?id=3341907}
\showURL{%
\tempurl}


\bibitem[\protect\citeauthoryear{Prana, Ford, Rastogi, Lo, Purandare, and
  Nagappan}{Prana et~al\mbox{.}}{2021}]%
        {prana2021geogenderdiversity}
\bibfield{author}{\bibinfo{person}{Gede Artha~Azriadi Prana},
  \bibinfo{person}{Denae Ford}, \bibinfo{person}{Ayushi Rastogi},
  \bibinfo{person}{David Lo}, \bibinfo{person}{Rahul Purandare}, {and}
  \bibinfo{person}{Nachiappan Nagappan}.} \bibinfo{year}{2021}\natexlab{}.
\newblock \showarticletitle{Including Everyone, Everywhere: Understanding
  Opportunities and Challenges of Geographic Gender-Inclusion in {OSS}}.
\newblock \bibinfo{journal}{\emph{IEEE Transactions on Software Engineering}}
  (\bibinfo{year}{2021}).
\newblock
\urldef\tempurl%
\url{https://doi.org/10.1109/TSE.2021.3092813}
\showDOI{\tempurl}
\newblock
\shownote{To appear}.


\bibitem[\protect\citeauthoryear{Qiu, Stewart, and Bartol}{Qiu
  et~al\mbox{.}}{2010}]%
        {qiu2010kdewomen}
\bibfield{author}{\bibinfo{person}{Yixin Qiu}, \bibinfo{person}{Katherine~J.
  Stewart}, {and} \bibinfo{person}{Kathryn~M. Bartol}.}
  \bibinfo{year}{2010}\natexlab{}.
\newblock \showarticletitle{Joining and Socialization in Open Source Women's
  Groups: An Exploratory Study of \emph{KDE-Women}}. In
  \bibinfo{booktitle}{\emph{6th International Conference on Open Source
  Systems, {OSS} 2010}} \emph{(\bibinfo{series}{{IFIP} Advances in Information
  and Communication Technology}, Vol.~\bibinfo{volume}{319})}.
  \bibinfo{publisher}{Springer}, \bibinfo{pages}{239--251}.
\newblock
\urldef\tempurl%
\url{https://doi.org/10.1007/978-3-642-13244-5\_19}
\showDOI{\tempurl}


\bibitem[\protect\citeauthoryear{Ralph, Baltes, Adisaputri, Torkar, Kovalenko,
  Kalinowski, Novielli, Yoo, Devroey, Tan, Zhou, Turhan, Hoda, Hata, Robles,
  Fard, and Alkadhi}{Ralph et~al\mbox{.}}{2020}]%
        {ralph2020pandemic}
\bibfield{author}{\bibinfo{person}{Paul Ralph}, \bibinfo{person}{Sebastian
  Baltes}, \bibinfo{person}{Gianisa Adisaputri}, \bibinfo{person}{Richard
  Torkar}, \bibinfo{person}{Vladimir Kovalenko}, \bibinfo{person}{Marcos
  Kalinowski}, \bibinfo{person}{Nicole Novielli}, \bibinfo{person}{Shin Yoo},
  \bibinfo{person}{Xavier Devroey}, \bibinfo{person}{Xin Tan},
  \bibinfo{person}{Minghui Zhou}, \bibinfo{person}{Burak Turhan},
  \bibinfo{person}{Rashina Hoda}, \bibinfo{person}{Hideaki Hata},
  \bibinfo{person}{Gregorio Robles}, \bibinfo{person}{Amin~Milani Fard}, {and}
  \bibinfo{person}{Rana Alkadhi}.} \bibinfo{year}{2020}\natexlab{}.
\newblock \showarticletitle{Pandemic programming}.
\newblock \bibinfo{journal}{\emph{Empir. Softw. Eng.}} \bibinfo{volume}{25},
  \bibinfo{number}{6} (\bibinfo{year}{2020}), \bibinfo{pages}{4927--4961}.
\newblock
\urldef\tempurl%
\url{https://doi.org/10.1007/s10664-020-09875-y}
\showDOI{\tempurl}


\bibitem[\protect\citeauthoryear{Rastogi}{Rastogi}{2016}]%
        {rastogi2016geobias}
\bibfield{author}{\bibinfo{person}{Ayushi Rastogi}.}
  \bibinfo{year}{2016}\natexlab{}.
\newblock \showarticletitle{Do biases related to geographical location
  influence work-related decisions in {GitHub}?}. In
  \bibinfo{booktitle}{\emph{Proceedings of the 38th International Conference on
  Software Engineering, {ICSE} 2016, Austin, TX, USA, May 14-22, 2016 -
  Companion Volume}}, \bibfield{editor}{\bibinfo{person}{Laura~K. Dillon},
  \bibinfo{person}{Willem Visser}, {and} \bibinfo{person}{Laurie~A. Williams}}
  (Eds.). \bibinfo{publisher}{{ACM}}, \bibinfo{pages}{665--667}.
\newblock
\urldef\tempurl%
\url{https://doi.org/10.1145/2889160.2891035}
\showDOI{\tempurl}


\bibitem[\protect\citeauthoryear{Reinking and Martin}{Reinking and
  Martin}{2018}]%
        {reinking2018genderstem}
\bibfield{author}{\bibinfo{person}{Anni Reinking} {and}
  \bibinfo{person}{Barbara Martin}.} \bibinfo{year}{2018}\natexlab{}.
\newblock \bibinfo{booktitle}{\emph{The gender gap in {STEM} fields: Theories,
  movements, and ideas to engage girls in {STEM}}}.
\newblock \bibinfo{type}{{T}echnical {R}eport}.
  \bibinfo{institution}{University of Alicante}.
\newblock


\bibitem[\protect\citeauthoryear{Robles, {Arjona Reina},
  Gonz{\'{a}}lez{-}Barahona, and Dom{\'{\i}}nguez}{Robles
  et~al\mbox{.}}{2016}]%
        {robles2016womeninfoss}
\bibfield{author}{\bibinfo{person}{Gregorio Robles}, \bibinfo{person}{Laura
  {Arjona Reina}}, \bibinfo{person}{Jes{\'{u}}s~M. Gonz{\'{a}}lez{-}Barahona},
  {and} \bibinfo{person}{Santiago~Due{\~{n}}as Dom{\'{\i}}nguez}.}
  \bibinfo{year}{2016}\natexlab{}.
\newblock \showarticletitle{Women in Free/Libre/Open Source Software: The
  Situation in the 2010s}. In \bibinfo{booktitle}{\emph{12th International
  Conference on Open Source Systems, {OSS} 2016}}
  \emph{(\bibinfo{series}{{IFIP} Advances in Information and Communication
  Technology}, Vol.~\bibinfo{volume}{472})}. \bibinfo{publisher}{Springer},
  \bibinfo{pages}{163--173}.
\newblock
\urldef\tempurl%
\url{https://doi.org/10.1007/978-3-319-39225-7\_13}
\showDOI{\tempurl}


\bibitem[\protect\citeauthoryear{Robles and Gonz{\'{a}}lez{-}Barahona}{Robles
  and Gonz{\'{a}}lez{-}Barahona}{2006}]%
        {robles2006devlocation}
\bibfield{author}{\bibinfo{person}{Gregorio Robles} {and}
  \bibinfo{person}{Jes{\'{u}}s~M. Gonz{\'{a}}lez{-}Barahona}.}
  \bibinfo{year}{2006}\natexlab{}.
\newblock \showarticletitle{Geographic location of developers at SourceForge}.
  In \bibinfo{booktitle}{\emph{Proceedings of the 2006 International Workshop
  on Mining Software Repositories, {MSR} 2006, Shanghai, China, May 22-23,
  2006}}, \bibfield{editor}{\bibinfo{person}{Stephan Diehl},
  \bibinfo{person}{Harald~C. Gall}, {and} \bibinfo{person}{Ahmed~E. Hassan}}
  (Eds.). \bibinfo{publisher}{{ACM}}, \bibinfo{pages}{144--150}.
\newblock
\urldef\tempurl%
\url{https://doi.org/10.1145/1137983.1138017}
\showDOI{\tempurl}


\bibitem[\protect\citeauthoryear{Rossi and Zacchiroli}{Rossi and
  Zacchiroli}{2022}]%
        {replication-package}
\bibfield{author}{\bibinfo{person}{Davide Rossi} {and} \bibinfo{person}{Stefano
  Zacchiroli}.} \bibinfo{year}{2022}\natexlab{}.
\newblock \bibinfo{booktitle}{\emph{Worldwide Gender Differences in Public Code
  Contributions - Replication Package}}.
\newblock
\urldef\tempurl%
\url{https://doi.org/10.5281/zenodo.6020475}
\showDOI{\tempurl}


\bibitem[\protect\citeauthoryear{Santamar{\'{\i}}a and
  Mihaljevic}{Santamar{\'{\i}}a and Mihaljevic}{2018}]%
        {santamaria2018genderapi}
\bibfield{author}{\bibinfo{person}{Luc{\'{\i}}a Santamar{\'{\i}}a} {and}
  \bibinfo{person}{Helena Mihaljevic}.} \bibinfo{year}{2018}\natexlab{}.
\newblock \showarticletitle{Comparison and benchmark of name-to-gender
  inference services}.
\newblock \bibinfo{journal}{\emph{PeerJ Computer Science}}  \bibinfo{volume}{4}
  (\bibinfo{year}{2018}), \bibinfo{pages}{e156}.
\newblock
\urldef\tempurl%
\url{https://doi.org/10.7717/peerj-cs.156}
\showDOI{\tempurl}


\bibitem[\protect\citeauthoryear{Terrell, Kofink, Middleton, Rainear,
  Murphy-Hill, Parnin, and Stallings}{Terrell et~al\mbox{.}}{2017}]%
        {terrell2017gender}
\bibfield{author}{\bibinfo{person}{Josh Terrell}, \bibinfo{person}{Andrew
  Kofink}, \bibinfo{person}{Justin Middleton}, \bibinfo{person}{Clarissa
  Rainear}, \bibinfo{person}{Emerson Murphy-Hill}, \bibinfo{person}{Chris
  Parnin}, {and} \bibinfo{person}{Jon Stallings}.}
  \bibinfo{year}{2017}\natexlab{}.
\newblock \showarticletitle{Gender differences and bias in open source: Pull
  request acceptance of women versus men}.
\newblock \bibinfo{journal}{\emph{PeerJ Computer Science}}  \bibinfo{volume}{3}
  (\bibinfo{year}{2017}), \bibinfo{pages}{e111}.
\newblock


\bibitem[\protect\citeauthoryear{Trinkenreich}{Trinkenreich}{2021}]%
        {trinkenreich2021pleasedontgo}
\bibfield{author}{\bibinfo{person}{Bianca Trinkenreich}.}
  \bibinfo{year}{2021}\natexlab{}.
\newblock \showarticletitle{Please Don't Go - {A} Comprehensive Approach to
  Increase Women's Participation in Open Source Software}. In
  \bibinfo{booktitle}{\emph{43rd {IEEE/ACM} International Conference on
  Software Engineering: Companion Proceedings, {ICSE} Companion 2021, Madrid,
  Spain, May 25-28, 2021}}. \bibinfo{publisher}{{IEEE}},
  \bibinfo{pages}{293--298}.
\newblock
\urldef\tempurl%
\url{https://doi.org/10.1109/ICSE-Companion52605.2021.00131}
\showDOI{\tempurl}


\bibitem[\protect\citeauthoryear{Trinkenreich, Guizani, Wiese, Sarma, and
  Steinmacher}{Trinkenreich et~al\mbox{.}}{2020}]%
        {trinkenreich2020hiddenfigures}
\bibfield{author}{\bibinfo{person}{Bianca Trinkenreich},
  \bibinfo{person}{Mariam Guizani}, \bibinfo{person}{Igor Wiese},
  \bibinfo{person}{Anita Sarma}, {and} \bibinfo{person}{Igor Steinmacher}.}
  \bibinfo{year}{2020}\natexlab{}.
\newblock \showarticletitle{Hidden Figures: Roles and Pathways of Successful
  {OSS} Contributors}.
\newblock \bibinfo{journal}{\emph{Proc. {ACM} Hum. Comput. Interact.}}
  \bibinfo{volume}{4}, \bibinfo{number}{{CSCW2}} (\bibinfo{year}{2020}),
  \bibinfo{pages}{180:1--180:22}.
\newblock
\urldef\tempurl%
\url{https://doi.org/10.1145/3415251}
\showDOI{\tempurl}


\bibitem[\protect\citeauthoryear{Vasilescu, Capiluppi, and
  Serebrenik}{Vasilescu et~al\mbox{.}}{2014}]%
        {vasilescu2014gender}
\bibfield{author}{\bibinfo{person}{Bogdan Vasilescu}, \bibinfo{person}{Andrea
  Capiluppi}, {and} \bibinfo{person}{Alexander Serebrenik}.}
  \bibinfo{year}{2014}\natexlab{}.
\newblock \showarticletitle{Gender, representation and online participation: A
  quantitative study}.
\newblock \bibinfo{journal}{\emph{Interacting with Computers}}
  \bibinfo{volume}{26}, \bibinfo{number}{5} (\bibinfo{year}{2014}),
  \bibinfo{pages}{488--511}.
\newblock


\bibitem[\protect\citeauthoryear{Vasilescu, Posnett, Ray, van~den Brand,
  Serebrenik, Devanbu, and Filkov}{Vasilescu et~al\mbox{.}}{2015}]%
        {vasilescu2015gender}
\bibfield{author}{\bibinfo{person}{Bogdan Vasilescu}, \bibinfo{person}{Daryl
  Posnett}, \bibinfo{person}{Baishakhi Ray}, \bibinfo{person}{Mark~GJ van~den
  Brand}, \bibinfo{person}{Alexander Serebrenik}, \bibinfo{person}{Premkumar
  Devanbu}, {and} \bibinfo{person}{Vladimir Filkov}.}
  \bibinfo{year}{2015}\natexlab{}.
\newblock \showarticletitle{Gender and tenure diversity in {GitHub} teams}. In
  \bibinfo{booktitle}{\emph{33rd annual {ACM} conference on human factors in
  computing systems, {CHI}'15}}. \bibinfo{pages}{3789--3798}.
\newblock


\bibitem[\protect\citeauthoryear{Wang and Degol}{Wang and Degol}{2017}]%
        {wang2017genderstem}
\bibfield{author}{\bibinfo{person}{Ming-Te Wang} {and}
  \bibinfo{person}{Jessica~L Degol}.} \bibinfo{year}{2017}\natexlab{}.
\newblock \showarticletitle{Gender gap in science, technology, engineering, and
  mathematics ({STEM}): Current knowledge, implications for practice, policy,
  and future directions}.
\newblock \bibinfo{journal}{\emph{Educational psychology review}}
  \bibinfo{volume}{29}, \bibinfo{number}{1} (\bibinfo{year}{2017}),
  \bibinfo{pages}{119--140}.
\newblock


\bibitem[\protect\citeauthoryear{Zacchiroli}{Zacchiroli}{2021}]%
        {zacchiroli2021gender}
\bibfield{author}{\bibinfo{person}{Stefano Zacchiroli}.}
  \bibinfo{year}{2021}\natexlab{}.
\newblock \showarticletitle{Gender Differences in Public Code Contributions:
  {A} 50-Year Perspective}.
\newblock \bibinfo{journal}{\emph{{IEEE} Softw.}} \bibinfo{volume}{38},
  \bibinfo{number}{2} (\bibinfo{year}{2021}), \bibinfo{pages}{45--50}.
\newblock
\urldef\tempurl%
\url{https://doi.org/10.1109/MS.2020.3038765}
\showDOI{\tempurl}

\end{thebibliography}
\end{document}